\documentclass[reqno,12pt]{amsart}
\usepackage{fullpage}
\usepackage{amsfonts}
\usepackage{amssymb}

\numberwithin{equation}{section}
\def\rvec{{\vec r}}
\def\vvec{{\vec v}}

\def\rhatvec{{\hat r}}
\def\thhatvec{{\hat\theta}}
\def\Ltens{{\mathbf L}}
\def\Lhattens{{\mathbf{\hat L}}}
\def\Thvec{{\mathbf{\hat \Theta}}}
\def\LRLvec{{\vec A}}

\def\L{\mathcal{L}}

\def\arctanh{{\rm arctanh}}

\def\X{{\bf X}}
\def\pr{{\rm pr}}
\def\ext{{\rm ext}}

\def\const{{\rm const.}}
\def\Rnum{\mathbb{R}}

\def\sgn{{\rm sgn}}
\def\spans{{\rm span}}

\def\soln{{\rm soln.}}
\def\eff{{\rm eff.}}
\def\eq{{\rm eq.}}

\def\min{{\rm min}}
\def\max{{\rm max}}

\def\parder#1#2{\frac{\partial{#1}}{\partial{#2}}}

\newtheorem{prop}{Proposition}
\newtheorem{thm}{Theorem}

\def\propref#1{Proposition~\ref{#1}}
\def\thmref#1{Theorem~\ref{#1}}

\def\Ref#1{Ref.\cite{#1}}
\def\secref#1{Sec.~\ref{#1}}

\def\ie/{i.e.}
\def\eg/{e.g.}
\def\com/{constant of motion}
\def\coms/{constants of motion}
\def\eom/{equations of motion}
\def\LRL/{Laplace-Runge-Lenz}
\def\const{\rm{const.}}

\begin{document}

\title{Some new aspects of first integrals\\
and symmetries for central force dynamics}

\author{
Stephen C. Anco$^1$,
Tyler Meadows$^{2,1}$
Vincent Pascuzzi$^{3,1}$\\
\\\lowercase{\scshape{
${}^1$
Department of Mathematics and Statistics, 
Brock University\\
St. Catharines, ON L2S3A1, Canada}} \\
\lowercase{\scshape{
${}^2$
Department of Mathematics and Statistics, 
McMaster University\\
Hamilton, ON L8S 4K1, Canada}}\\
\lowercase{\scshape{
${}^3$
Department of Physics, 
University of Toronto\\
Toronto, ON M5S 1A7, Canada }}
}

\begin{abstract}
For the general central force \eom/ in $n>1$ dimensions, 
a complete set of $2n$ first integrals is derived in an explicit algorithmic way
without the use of dynamical symmetries or Noether's theorem. 
The derivation uses the polar formulation of the \eom/
and yields energy, angular momentum, a generalized \LRL/ vector, 
and a temporal quantity involving the time variable explicitly. 
A variant of the general \LRL/ vector, 
which generalizes Hamilton's eccentricity vector, is also obtained. 
The physical meaning of the general \LRL/ vector, its variant, 
and the temporal quantity are discussed for general central forces. 
Their properties are compared for precessing bounded trajectories 
versus non-precessing bounded trajectories, as well as unbounded trajectories,
by considering an inverse-square force (Kepler problem)
and a cubically perturbed inverse-square force 
(Newtonian revolving orbit problem). 
\end{abstract}

\maketitle

\section{\large Introduction}
\label{intro}

Classical central force dynamics have been long studied from both 
physical and mathematical viewpoints. 
On one hand, many basic physical models are described by central forces,
\eg/ planetary motion and Coulomb scattering, 
which have inverse-square radial forces;
vibrations of atoms in a crystal,
which have linear radial forces;
interactions between a pair of neutral atoms or molecules,
which have complicated nonlinear radial forces. 
On the other hand, 
the general \eom/ for central forces have a rich mathematical structure
and are important examples of completely integrable Hamiltonian systems
that possess the maximal number of \coms/ in involution. 

The \coms/ for central force dynamics in three spatial dimensions 
are well-known to consist of 
the energy, the angular momentum vector, and an additional vector \cite{Fra,Per}
which is a generalization of the usual \LRL/ vector 
known for inverse-square central forces \cite{GolPooSaf}. 
These \coms/ comprise altogether 7 first integrals of the \eom/
for a general central force. 
Only 5 of these first integrals are independent, 
since the generalized \LRL/ vector is orthogonal to the angular momentum vector,
while its magnitude can be expressed in terms of 
the energy and the magnitude of angular momentum,
so thus there are two relations among the 7 first integrals. 
The derivation of energy and angular momentum can be done in a simple way 
through Noether's theorem \cite{GolPooSaf} based on symmetries of 
the Lagrangian formulation of the \eom/.
Noether's theorem (in its most general form \cite{BluAnc,Olv}) 
shows that every group of symmetry transformations
under which the Lagrangian is invariant (to within a total derivative term) 
gives rise to a first integral of the \eom/,
and conversely, every first integral 
arises from a group of symmetry transformations of the Lagrangian. 
Energy is given by the group of time-translations, 
and the components of angular momentum are given by the $SO(3)$ group of rotations,
where these symmetry groups each act as point transformations
on the position variables and the time variable. 
The components of the \LRL/ vector, in contrast, 
arise from a hidden $SO(3)$ group of symmetry transformations 
which are not point transformations but instead turn out to be dynamical (first-order) symmetries 
when acting on the position and time variables. 
Indeed, 
a simple, explicit formulation of these hidden symmetry group transformations
is hard to find in the literature \cite{Rog} 
and typically only the generators are presented \cite{Fra,Lev}. 
Moreover, 
while the set of all point symmetries of the central force \eom/ is finite-dimensional, 
the set of all dynamical symmetries is infinite-dimensional 
since a general first-order symmetry will necessarily involve arbitrary functions of all of the first integrals \cite{BluAnc}. 
(In particular, for any dynamical system, 
multiplication of any symmetry by any function of first integrals automatically yields a symmetry, since first integrals are constants 
for all solutions of the system.)
Therefore, 
any direct calculation of the hidden $SO(3)$ symmetry group of the central force \eom/ starting from the determining equations for dynamical symmetries 
will require, at some step, the calculation of all first integrals, 
which undercuts the entire approach of using Noether's theorem to obtain the \LRL/ vector. 
See \Ref{PriEli,GodPri,Nuc} for some alternative approaches using symmetry ideas. 

The purpose of the present paper is to take a fresh, comprehensive look
at how to derive {\em all} of the first integrals 
for the general central force \eom/, 
without the use of dynamical symmetries or Noether's theorem. 
This problem will be studied in an arbitrary number of dimensions $n>1$.
Several interesting main new results will be obtained, which include 
deriving a general $n$-dimensional \LRL/ vector as a first integral 
in two explicit algorithmic ways, 
and showing a complete set of $2n$ first integrals is generated from 
this general \LRL/ vector, energy, angular momentum, 
and a temporal quantity involving the time variable explicitly. 
This temporal quantity provides a first integral that is not a \com/.
A new variant of the general \LRL/ vector will also be introduced,
which arises naturally from these derivations and provides a generalization of
Hamilton's eccentricity vector. 
The physical meaning of the general \LRL/ vector, its variant, 
and the temporal quantity 
will be discussed for general central forces,
and their properties will be compared for precessing bounded trajectories 
versus non-precessing bounded trajectories, as well as unbounded trajectories. 

In general, 
a {\em \com/} of $n$-dimensional dynamical \eom/
with position variable $\rvec$ and time $t$
is a function $C$ of $\rvec$ and $d\rvec/dt =\vvec$
whose time derivative vanishes 
\begin{equation}\label{ndim-com}
\frac{dC}{dt}(\rvec(t),\vvec(t))\big|_\soln =0
\end{equation}
for all solutions $\rvec(t)$ of the \eom/.
Similarly, 
a {\em first integral} is a function $I$ of $\rvec$, $\vvec$, and $t$
such that 
\begin{equation}\label{ndim-1stintegr}
\frac{dI}{dt}(t,\rvec(t),\vvec(t))\big|_\soln =0
\end{equation}
holds for all solutions $\rvec(t)$. 
Note that if a first integral does not involve $t$ explicitly, 
then it is a \com/. 
Because dynamical \eom/ in $n$ dimensions 
comprise a set of $n$ second-order differential equations for $\rvec(t)$, 
the number of functionally-independent first integrals is equal to $2n$,
while the number of functionally-independent \coms/ is equal to $2n-1$. 

Our starting point is the well-known fact that \cite{GolPooSaf}
every solution $\rvec(t)$ of the central force \eom/ in $n>1$ dimensions 
\begin{equation}\label{ndim-eom}
m\frac{d^2\rvec}{dt^2} = \frac{F(|\rvec|)}{|\rvec|} \rvec
\end{equation}
lies in a $2$-dimensional plane that is spanned by the initial position vector
$\rvec(0)$ and the initial velocity vector $\vvec(0)$. 
Thus the dynamics can be described by polar variables $r,\theta$ 
in this time-independent plane.
The reduced \eom/
\begin{equation}\label{polar-eom}
\frac{d^2r}{dt^2} = \omega^2 r + m^{-1}F(r) ,
\quad
\frac{d^2\theta}{dt^2} = -2 \omega v/r
\end{equation}
comprise a pair of coupled second-order nonlinear differential equations 
for the dynamical variables $r(t)$, $\theta(t)$,
where $v=dr/dt$ is the radial speed
and $\omega= d\theta/dt$ is the angular speed.
A first integral of the polar \eom/ \eqref{polar-eom} is a function 
$I(t,r,\theta,v,\omega)$ satisfying 
\begin{equation}\label{polar-I}
\frac{dI}{dt}(t,r(t),\theta(t),v(t),\omega(t))\big|_\soln
= I_t + vI_r + \omega I_\theta +(\omega^2 r + m^{-1} F(r)) I_v -(2\omega v/r)I_\omega
=0 . 
\end{equation}
This is the {\em determining equation} for all first integrals of these \eom/.

In \secref{prelim}, 
we will briefly review some aspects of this reduction of the $n$-dimensional problem, including the Lagrangian formulation of the polar \eom/. 
We also review the point symmetries of the polar \eom/ and show that for a general central force these symmetries comprise only time-translations and rotations.

In \secref{derivation}, 
we directly derive the general solution of the determining equation \eqref{polar-I} for polar first integrals, 
which yields four functionally-independent first integrals. 
Two of these first integrals are 
the energy $E$ and the scalar angular momentum $L$. 
The third first integral is an angular quantity $\Theta$ 
which corresponds to the angular direction of a general \LRL/ vector 
in the plane of motion. 
These three first integrals $E,L,\Theta$ are \coms/. 
The fourth first integral is a temporal quantity $T$ 
which involves $t$ explicitly. 
It is related to the angular quantity by the property that
$t=T$ when $\theta(t)=\Theta$, for all non-circular solutions $r(t)\neq\const$.
These two first integrals are not widely known or explicitly discussed 
in much of the literature. 
We conclude \secref{derivation} with a detailed discussion of 
some general aspects of the quantities $\Theta$ and $T$, 
especially their connection to 
apsides (turning points) on trajectories $(r(t),\theta(t))$ 
determined by solutions of the polar \eom/ \eqref{polar-eom}. 
We also discuss a variant of $\Theta$ which is connected to other distinguished points on trajectories. 

Next, in \secref{examples}, 
we illustrate the physical meaning of $\Theta$ and $T$ 
for two important examples of central force dynamics:
(1) inverse-square force $F=-k/r^2$, \eg/ Kepler problem;
(2) cubically perturbed inverse-square force $F=-k/r^2 -\kappa/r^3$, 
\eg/ Newtonian revolving orbit problem 
(which has precessing trajectories). 

In \secref{noether}, 
we discuss Noether's theorem for the polar \eom/,
and we use it to derive the symmetries corresponding to the first integrals. 
For the $\Theta$ and $T$ first integrals,
these symmetries are found to be dynamical (first-order) symmetries 
which involve both $v$ and $\omega$, 
while for the $L$ and $E$ first integrals, 
the symmetries consist of time-translations and polar rotations,
which are point symmetries. 
We show that this entire set of symmetries generates an abelian algebra
(namely, all of the symmetries commute with each other). 
We derive the corresponding groups of transformations in an explicit form
and work out how each group acts on all of the first integrals. 
In particular, 
these symmetry groups are shown to act in a simple way
as point transformations on $(t,r,\theta,L,E)$. 
We use this new result to formulate a novel method employing extended point symmetries
to derive the first integrals $\Theta$ and $T$ as well as the hidden dynamical symmetry group connected with them. 

In \secref{ndim},
we express the first integrals $E$, $L$, $\Theta$, $T$ 
in an $n$-dimensional geometric form in terms of the position vector $\rvec$ and velocity vector $\vvec$ in $\Rnum^n$. 
Through this geometrical correspondence,
$L$ is shown to yield an antisymmetric angular momentum tensor in the plane of motion,
and $\Theta$ is shown to yield a directional unit vector in the plane of motion.
We discus the general properties of this unit vector 
and show how it naturally gives rise to a general \LRL/ vector and a variant vector. 

In \secref{LRLexamples}, 
we write out the general \LRL/ vector and its variant in $n$ dimensions 
for the two examples of central force dynamics considered previously,
and we look at the physical and geometrical properties of the resulting $n$-dimensional vectors.

Finally, in \secref{remarks},
we make some concluding remarks and outline avenues for future work.

\section{Preliminaries}
\label{prelim}

The central force \eom/ \eqref{ndim-eom} in $\Rnum^n$ 
can be expressed equivalently as a first-order system 
\begin{equation}\label{ndim-sys}
\frac{d\rvec}{dt} = \vvec, 
\quad
\frac{d\vvec}{dt} = m^{-1}|\rvec|^{-1} F(|\rvec|)\rvec
\end{equation}
for position $\rvec$ and velocity $\vvec$ when $n>1$. 
If these two vectors are collinear for all time $t$ 
in a solution $(\rvec(t),\vvec(t))$,
then the position vector $\rvec(t)$ is clearly confined to 
a straight line in $\Rnum^n$. 
In this case the motion trivially lies in any time-independent plane 
that contains this line. 
If instead the vectors $\rvec$ and $\vvec$ are not collinear 
at some time $t$ in a solution, 
then $\spans(\rvec,\vvec)$ at time $t$ 
defines a $2$-dimensional plane in $\Rnum^n$.
To see that this plane is time-independent, 
we note that the \eom/ \eqref{ndim-sys} yield 
$\spans(d\rvec/dt,d\vvec/dt) =\spans(\vvec,\lambda\rvec) 
= \spans(\vvec,\rvec)$, 
where $\lambda=F(|\rvec|)|\rvec|^{-1}$. 
Hence the span is time-independent, 
which immediately implies that the motion lies in a time-independent plane 
in $\Rnum^n$ spanned by the initial position $\rvec(0)$ and initial velocity $\vvec(0)$. 

Let 
\begin{equation}\label{plane-basisvecs}
\{\hat e_1, \hat e_2\}
\end{equation}
be any fixed (time-independent) orthonormal basis for the plane of motion. 
Then we have 
\begin{equation}\label{polar-vecs}
\rvec = (r\cos\theta)\hat e_1 + (r\sin\theta)\hat e_2,
\quad
\vvec = (v\cos\theta-\omega r\sin\theta)\hat e_1 + (v\sin\theta+\omega r\cos\theta)\hat e_2 
\end{equation}
where $r$ and $\theta$ are polar variables,
and $v=dr/dt$ and $\omega=d\theta/dt$ are their time derivatives. 
Substitution of this polar representation into the \eom/ \eqref{ndim-sys} 
yields the polar \eom/ \eqref{polar-eom} 
in which $r$ and $\theta$ are the dynamical variables. 
This reduces the number of degrees of freedom from $n>1$ to $2$. 

The polar \eom/ \eqref{polar-eom} arise from the well-known Lagrangian 
\begin{equation}\label{polar-Lagr}
\L = \tfrac{1}{2}m(v^2 +r^2\omega^2) - U(r)
\end{equation}
where $U(r)$ is the radial potential determined (up to an additive constant)
by the central force
\begin{equation}\label{potential}
F(r) = -U'(r) . 
\end{equation}
In particular, we have 
\begin{equation}\label{polar-Lagr-varders}
\frac{\delta\L}{\delta r} = 
m\omega^2 r + F(r) -m\frac{d^2r}{dt^2}, 
\quad
\frac{\delta\L}{\delta\theta} = 
-2m \omega vr - mr^2\frac{d^2\theta}{dt^2} . 
\end{equation}

Hereafter, we will put $m=1$ without loss of generality
(via rescaling the physical units of the dynamical variables). 

In polar variables, 
a {\em point transformation} is an invertible mapping of 
the coordinate space $(t,r,\theta)$ into itself. 
We will be interested in Lie groups of point transformations. 
A one-dimensional Lie group of point transformations, 
with group parameter $\epsilon$, consists of 
\begin{equation}\label{polar-pointtransf}
t\rightarrow t_{(\epsilon)}(t,r,\theta),
\quad
r\rightarrow r_{(\epsilon)}(t,r,\theta),
\quad
\theta\rightarrow \theta_{(\epsilon)}(t,r,\theta),
\quad
\infty<\epsilon<\infty
\end{equation}
such that $\epsilon=0$ yields the identity transformation. 
The generator of the group is an infinitesimal transformation
\begin{equation}\label{polar-infpointtransf}
\delta t = \parder{t_{(\epsilon)}}{\epsilon}\Big|_{\epsilon=0}
=\tau(t,r,\theta),
\quad
\delta r = \parder{r_{(\epsilon)}}{\epsilon}\Big|_{\epsilon=0}
=\xi(t,r,\theta),
\quad
\delta \theta = \parder{\theta_{(\epsilon)}}{\epsilon}\Big|_{\epsilon=0}
=\psi(t,r,\theta)
\end{equation}
which can be viewed as defining a vector field 
\begin{equation}\label{polar-X}
\X= \tau\partial_t + \xi\partial_r + \psi\partial_\theta
\end{equation}
on the space of variables $(t,r,\theta)$. 
From this infinitesimal generator, 
the point transformations \eqref{polar-pointtransf} 
can be obtained by exponentiation of the vector field \eqref{polar-X}. 

Point transformations have a natural action on functions $(r(t),\theta(t))$.
In infinitesimal form, this action is given by 
\begin{equation}
\delta r(t) = \xi(t,r(t),\theta(t)) - v(t)\tau(t,r(t),\theta(t)),
\quad
\delta \theta(t) = \psi(t,r(t),\theta(t)) - \omega(t)\tau(t,r(t),\theta(t))
\end{equation}
where $v(t)=dr(t)/dt$ and $\omega(t)=d\theta(t)/dt$. 
The corresponding generator can be expressed as a vector field
\begin{equation}\label{polar-Xhat}
\hat\X= P^r\partial_r + P^\theta\partial_\theta, 
\quad
P^r= \xi-v\tau,
\quad
P^\theta= \psi-\omega\tau . 
\end{equation}
This vector field is called the {\em characteristic form} \cite{BluAnc,Olv}
associated to the generator \eqref{polar-X}. 

Both vector fields \eqref{polar-Xhat} and \eqref{polar-X} can be prolonged
to act on time-derivatives of the polar variables. 
The prolongation of the vector field \eqref{polar-Xhat} is very simple,
\begin{equation}\label{polar-prXhat}
\pr\hat\X= \hat\X + \frac{dP^r}{dt}\partial_v + \frac{dP^\theta}{dt}\partial_\omega + \cdots
\end{equation}
while the prolongation of the vector field \eqref{polar-X} takes the related form 
\begin{equation}\label{polar-prXrel}
\pr\X= \pr\hat\X + \tau\frac{d}{dt}
\end{equation}
where the total time derivative is viewed as a vector field 
\begin{equation}\label{polar-Dt}
\frac{d}{dt} = \partial_t + v\partial_r + \omega\partial_\theta + \cdots
\end{equation}
by the chain rule. 
Note these prolonged vector fields are defined in the coordinate space 
$(t,r,\theta,v,\omega,\ldots)$, 
which is called the jet space of the polar \eom/.

A {\em point symmetry} of the polar \eom/ \eqref{polar-eom} 
is a Lie group of point transformations \eqref{polar-pointtransf} that leaves
invariant the solution space of the equations. 
Infinitesimal invariance of the solution space is expressed in terms of 
the prolonged generator \eqref{polar-X} of the transformation 
acting on the polar \eom/ by 
\begin{equation}\label{polar-X-eom}
\pr\X\left( \frac{d^2r}{dt^2} - \omega^2 r -F(r) \right)\Big|_{\soln} =0,
\quad
\pr\X\left( \frac{d^2\theta}{dt^2} +2 \omega v/r \right)\Big|_{\soln} =0
\end{equation}
which is required to hold for all solutions of the equations. 
This invariance condition can be expressed more simply 
through the characteristic generator \eqref{polar-Xhat}, 
which has an equivalent action on the \eom/ 
because of the relation \eqref{polar-Dt}. 
In particular, 
since a total time-derivative of the \eom/ necessarily vanishes 
on solutions of the equations, 
the invariance condition becomes 
\begin{equation}\label{polar-Xhat-eom}
\begin{aligned}
& \pr\hat\X\left( \frac{d^2r}{dt^2} - \omega^2 r -F(r) \right)\Big|_{\soln} 
= \frac{d^2P^r}{dt^2} - \omega^2P^r -2\omega r \frac{dP^\theta}{dt} -F'(r) P^r 
=0
\\
& \pr\hat\X\left( \frac{d^2\theta}{dt^2} +2 \omega v/r \right)\Big|_{\soln}
= \frac{d^2P^\theta}{dt^2} +2\frac{dP^\theta}{dt} vr^{-1} +2\omega r^{-1} \frac{dP^r}{dt} -2\omega vr^{-2} P^r
=0
\end{aligned}
\end{equation}
where $d^2r/dt^2$, $d^2\theta/dt^2$, and all higher-order time derivatives 
are eliminated through use of the \eom/ \eqref{polar-eom}. 
Thus, 
the generator of a point transformation \eqref{polar-X} 
will be an infinitesimal symmetry of the polar \eom/ iff 
the associated characteristic generator \eqref{polar-Xhat} satisfies 
the invariance condition \eqref{polar-Xhat-eom}. 
For this reason the two generators \eqref{polar-X} and \eqref{polar-Xhat} 
are commonly referred to as equivalent symmetry vector fields,
and the invariance condition \eqref{polar-Xhat-eom} is called 
the {\em determining equation} \cite{BluAnc,Olv} for point symmetries. 

The determining equation \eqref{polar-Xhat-eom} is 
a straightforward linear partial differential equation 
to be solved for the infinitesimal symmetry components 
$\tau(t,r,\theta)$, $\xi(t,r,\theta)$, and $\psi(t,r,\theta)$. 
In particular, 
this equation will split with respect to $v$ and $\omega$ in the jet space $(t,r,\theta,v,\omega)$, 
yielding a linear overdetermined system on the functions 
$\tau(t,r,\theta)$, $\xi(t,r,\theta)$, and $\psi(t,r,\theta)$. 
If we regard the force $F(r)$ as an additional unknown, 
then we can also determine any infinitesimal point symmetries that hold only 
for special force expressions $F(r)$. 
In all cases, the set of admitted infinitesimal point symmetries 
forms a Lie algebra.

By a direct calculation (using Maple), 
we obtain the following result. 

\begin{thm}\label{polar-pointsymm}
The infinitesimal point symmetries \eqref{polar-X} admitted by 
the polar \eom/ \eqref{polar-eom} with a nonlinear force $F(r)$
are spanned by the generators:
\begin{align}
&\begin{aligned}
(\rm a)\quad
&\text{general } F(r)\\
& 
\X_1=\partial_\theta
\quad
\X_2=\partial_t,
\end{aligned}
\\
&\begin{aligned}
(\rm b)\quad
& F(r)=kr^p, 
\quad 
p\neq 1\\
& 
\X_1=\partial_\theta,
\quad
\X_2=\partial_t,
\quad
\X_3=t\partial_t +\frac{2}{1-p}r\partial _r
\end{aligned}
\\
&\begin{aligned}
(\rm c)\quad
& F(r)=kr +\tilde kr^{-3},
\quad 
k\neq 0, \tilde k\neq 0\\
& 
\X_1=\partial_\theta,
\quad
\X_2=\partial_t,
\quad
\X_4=\exp(2k^{1/2}t)\left(\partial_t +k^{1/2}r\partial_r\right)
\end{aligned}
\end{align}
The point symmetry transformations generated by $X_1$, $X_2$, $X_3$, and $X_4$
are, respectively, given by 
\begin{align}
&({\rm 1})\quad
\theta\rightarrow \theta+\epsilon
\qquad \text{polar rotation}
\label{polar-rot}\\
&({\rm 2})\quad
t\rightarrow t+\epsilon
\qquad \text{time-translation}
\label{polar-trans}\\
&({\rm 3})\quad
t\rightarrow \exp(\epsilon)t,
\quad
r\rightarrow \exp\left(2\epsilon/(1-p)\right)r
\qquad \text{scaling}
\\
&({\rm 4})\quad
t\rightarrow t+k^{-1/2}\ln\big(\left(1-2k^{1/2}\epsilon\exp(2k^{1/2}t)\right)^{-1/2}\big),
\quad
r\rightarrow r\left(1-2k^{1/2}\epsilon\exp(2k^{1/2}t)\right)^{-1/2}
\nonumber\\
&\;\qquad \text{time-dependent dilation}
\end{align}
\end{thm}

Notice that only two point symmetries are admitted 
for a general central force $F(r)$, 
whereas the \eom/ possess four functionally independent first integrals. 
Hence, point symmetries of the \eom/ are not rich enough to capture 
all of the first integrals. 

An important generalization of point symmetries 
is provided by dynamical symmetries. 
A {\em dynamical symmetry} \cite{Ste} of the polar \eom/ \eqref{polar-eom} 
consists of an infinitesimal transformation of the general form \eqref{polar-X}
in which the components $\tau$, $\xi$, $\psi$ are allowed to depend on 
the radial speed $v$ and angular speed $\omega$ 
(in addition to the variables $t,r,\theta$) 
such that the invariance condition \eqref{polar-X-eom}
holds for all solutions of the \eom/. 
As in the case of point symmetries, 
it is simpler to work with the equivalent characteristic generator \eqref{polar-Xhat} for dynamical symmetries,
which satisfies the determining equation \eqref{polar-Xhat-eom}. 

Similarly to point symmetries, 
dynamical symmetries can be exponentiated to obtain 
a group of transformations acting on solutions of the polar \eom/. 
These transformations take the form of point transformations 
acting on the dynamical variables $(r(t),\theta(t),v(t),\omega(t))$
given by any solution $(r(t),\theta(t))$, 
but they cannot be extended off solutions 
to act on the coordinate space $(t,r,\theta,v,\omega)$. 

Moreover, in further contrast to point symmetries, 
the determining equation \eqref{polar-Xhat-eom} for dynamical symmetries 
no longer splits with respect to $v$ and $\omega$ in the jet space $(t,r,\theta,v,\omega)$,
since these variables now appear in the dynamical symmetry components 
$\tau(t,r,\theta,v,\omega)$, $\xi(t,r,\theta,v,\omega)$, and $\psi(t,r,\theta,v,\omega)$. 
This means that the determining equation is 
a coupled pair of linear partial differential equations,
and in general we cannot solve these equations without already knowing
how to integrate the polar \eom/ themselves. 
In particular, 
the general solution of the dynamical symmetry determining equation 
will involve arbitrary functions of all first integrals of the polar \eom/. 

Therefore, 
it will be more useful to derive the first integrals directly from the \eom/, 
rather than use a symmetry approach. 
Nevertheless, dynamical symmetries have a direct connection to first integrals
through Noether's theorem,
which can be used to find the specific dynamical symmetries that correspond to 
the first integrals in \thmref{allI}.

\section{Derivation and properties of polar first integrals}
\label{derivation}

For the polar \eom/ \eqref{polar-eom} of central force dynamics,
all first integrals are functions $I(t,r,\theta,v,\omega)$  
determined by equation \eqref{polar-I}. 
With $m=1$ and $F(r)=-U'(r)$, 
this determining equation becomes 
\begin{equation}\label{deteq}
\frac{dI}{dt}\Big|_\soln
= I_t + vI_r + \omega I_\theta +(\omega^2 r -U'(r)) I_v -(2\omega vr^{-1})I_\omega
=0
\end{equation}
which is a linear first-order partial differential equation. 
Its general solution can be obtained through the method of characteristics 
\cite{Joh}
by integrating the system of differential equations 
\begin{equation}
\frac{dt}{1} 
= \frac{dr}{v} 
= \frac{d\theta}{\omega} 
= \frac{dv}{\omega^2r-U'(r)} 
= \frac{d\omega}{-2\omega v r^{-1}}
=\frac{dI}{0} . 
\end{equation}
This system can be arranged in a triangular form 
\begin{align}
&
\frac{d\omega}{dr} = \frac{-2\omega}{r}
\label{omegaeq}\\
&
\frac{dv}{dr} = \frac{\omega^2r-U'(r)}{v}
\label{veq}\\
&
\frac{d\theta}{dr} = \frac{\omega}{v}
\label{thetateq}\\
&
\frac{dt}{dr} = \frac{1}{v}
\label{teq}\\
&
\frac{dI}{dr} = 0
\label{Ieq}
\end{align}
whereby the successive integration of 
these differential equations \eqref{omegaeq}--\eqref{teq} 
is achieved by separation of variables. 
Each constant of integration will then be a particular first integral 
satisfying the determining equation \eqref{deteq}. 

Separating variables in equation \eqref{omegaeq}, 
we obtain the first integral 
\begin{equation}\label{I1}
I_1 = \omega r^2 . 
\end{equation}
We can now express $\omega$ in terms of $I_1$ and $r$:
\begin{equation}
\omega = I_1 r^{-2} . 
\end{equation}
Then equation \eqref{veq} is separable
\begin{equation}
\frac{dv}{dr} = \frac{I_1{}^2 r^{-3}-U'(r)}{v} . 
\end{equation}
Hence we obtain a second first integral 
\begin{equation}\label{I2}
I_2= \tfrac{1}{2}v^2 + \tfrac{1}{2} I_1^2r^{-2} +U(r) . 
\end{equation}
We can next express the magnitude of $v$ in terms of $I_1$, $I_2$, and $r$:
\begin{equation}
|v|=\sqrt{ 2(I_2 -U(r)) -I_1{}^2 r^{-2}} . 
\end{equation}
Both equations \eqref{thetateq} and \eqref{teq} now become separable, 
\begin{equation}
\frac{d\theta}{dr} = \frac{\sgn(v)I_1}{r^2\sqrt{ 2(I_2 -U(r))-I_1{}^2 r^{-2}}}
\end{equation}
and
\begin{equation}
\frac{dt}{dr} = \frac{\sgn(v)}{\sqrt{ 2(I_2-U(r))- I_1{}^2 r^{-2}}} . 
\end{equation}
Hence we obtain two more first integrals
\begin{equation}\label{I3}
I_3 = \theta - I_1\int \frac{\sgn(v)}{\sqrt{ 2(I_2-U(r))r^4 -I_1{}^2 r^2}}\,dr
\equiv \Theta 
\end{equation}
and
\begin{equation}\label{I4}
I_4 = t- \int \frac{\sgn(v)}{\sqrt{ 2(I_2-U(r)) -I_1{}^2 r^{-2}}}\,dr 
\equiv T . 
\end{equation}
Finally, the remaining differential equation \eqref{Ieq} shows that 
$I$ is an arbitrary function of the previous four first integrals 
\eqref{I1}, \eqref{I2}, \eqref{I3}, \eqref{I4}. 
The following result is now immediate. 

\begin{prop}
The general solution of the determining equation \eqref{deteq}
for first integrals of the polar \eom/ \eqref{polar-eom} is given by 
$I=f(I_1,I_2,I_3,I_4)$
where $f$ is an arbitrary differentiable function 
and where $I_1$, $I_2$, $I_3$, $I_4$ have physical units of 
angular momentum, energy, radians, and time, respectively. 
\end{prop}

Note that these first integrals $I_1$, $I_2$, $I_3$, $I_4$ are functionally independent 
because they each have different physical units. 
Hence, 
$I_1$, $I_2$, $I_3$, $I_4$ provide the complete quadrature of the polar \eom/.

\subsection{Normalization (``zero-point'' values)}

Any first integral $I$ remains conserved if an arbitrary constant is added to it.
This freedom represents the choice of a ``zero-point'' value 
for the physical quantity defined by the first integral. 
In particular, we can write 
\begin{align}
& I_1 = L+L_0 
\label{I1=L}\\
& I_2 = E+E_0 
\label{I2=E}\\
& I_3 = \Theta+\Theta_0 
\label{I3=Theta}\\
& I_4 = T+T_0
\label{I4=T}
\end{align}
where $L_0$, $E_0$, $\Theta_0$, $T_0$ will denote the zero-point constants,
and $L$, $E$, $\Theta$, $T$ will be the normalized physical quantities,
which have units of angular momentum, energy, radians, and time, respectively.

The zero-point constants in the first integrals need to be specified by 
some additional considerations such that the resulting quantities 
$L$, $E$, $\Theta$, $T$ are physically meaningful and mathematically well-defined
for all solutions of the polar \eom/ \eqref{polar-eom}. 
It is clear how we can do this for 
the angular momentum constant $L_0$ and the energy constant $E_0$,
by adopting the standard Newtonian expressions for 
the physical angular momentum $L$ and the physical energy $E$ 
for motion under a central force. 
But since the physical interpretation of $\Theta$ and $T$ is not obvious
just from their expressions, 
we will instead use a different argument to determine all four constants 
$L_0$, $E_0$, $\Theta_0$, $T_0$ based on general properties of 
the effective potential for the \eom/. 
Normalization of $\Theta$ is often overlooked in the literature yet is crucial 
for understanding its relationship with the \LRL/ vector. 

From the first integrals \eqref{I1} and \eqref{I2}, 
written in the respective forms \eqref{I1=L} and \eqref{I2=E}, 
the effective potential is defined by 
\begin{equation}
U_\eff(r) = \tfrac{1}{2} (L+L_0)^2 r^{-2} + U(r) -E_0 . 
\end{equation}
This potential determines the types of trajectories 
admitted for solutions of the polar \eom/ \eqref{polar-eom}. 
In all cases of physical interest, 
we may suppose that the effective potential has at least one 
{\em equilibrium point}, $r=r_\eq$, 
defined by the condition that the central force $-U'(r)$ vanishes at $r=r_\eq$
(which can include $r=0$ or $r=\infty$). 
This condition coincides with the effective force $-U_\eff'(r)$ being zero 
when both the radial and angular speeds are zero.  
Thus, the set of equilibrium points is given by the roots of the equation 
\begin{equation}\label{equilpoint}
0= -U'(r_\eq)= -U_\eff'(r_\eq)\big|_{L+L_0=0} . 
\end{equation}
The physical meaning of an equilibrium point is that it corresponds to 
a static solution of the polar \eom/ \eqref{polar-eom}, 
with $r=r_\eq=\const$ (and $\theta=\const$).

A natural condition to determine $L_0$ and $E_0$ is that both
$L|_{r=r_\eq}$ and $E|_{r=r_\eq}$ must be zero 
when these two first integrals are evaluated for static solutions of 
the polar \eom/ \eqref{polar-eom} 
given by an equilibrium point \eqref{equilpoint} away from the origin, 
$r=r_\eq\neq 0$. 
Since a static solution has $\omega=v=0$,
the condition $L|_{r=r_\eq}=0$ directly implies 
\begin{equation}
L_0=0
\end{equation}
in the first integral \eqref{I1=L},
while the condition $E|_{r=r_\eq}=0$ together with $L|_{r=r_\eq}=L_0=0$ 
yields
\begin{equation}
E_0=U(r_\eq) 
\end{equation}
in the first integral \eqref{I2=E}. 
This gives 
\begin{equation}\label{L}
L = \omega r^2 
\end{equation}
and 
\begin{equation}\label{E}
E= \tfrac{1}{2}v^2 + \tfrac{1}{2} L^2r^{-2} +U(r) -U(r_\eq)
\end{equation}
which are the usual Newtonian definitions of 
physical angular momentum and physical energy 
for central force dynamics. 
Note the effective potential now simplifies:
\begin{equation}\label{Ueff}
U_\eff(r) = \tfrac{1}{2} L^2 r^{-2} + U(r) -U(r_\eq) . 
\end{equation}

This equilibrium point argument, however, 
will not extend directly to $\Theta_0$ and $T_0$, 
because $\Theta$ and $T$ can have any values at an equilibrium point $r=r_\eq$. 
To determine $\Theta_0$ and $T_0$, 
we will consider, more generally, 
distinguished points of the effective potential other than equilibrium points. 
Two natural distinguished points are inertial points and turning points, 
as these points are defined solely in terms of $U_\eff(r)$ and $E$.

An {\em inertial point} is a finite radial value $r=r^*$ 
at which the effective force vanishes. 
The set of all inertial points is thus given by the roots (if any) of 
the effective force equation
\begin{equation}\label{inertialpoint}
0= -U_\eff'(r^*) = L^2 r^*{}^{-3} - U'(r^*), 
\quad
0\leq r^*<\infty . 
\end{equation}
This is a generalization of the equation \eqref{equilpoint} 
defining an equilibrium point if$L\neq 0$. 
Hence, a solution of the polar \eom/ \eqref{polar-eom} possesses an inertial point
if (and only if) both $dv/dt=0$ and $\omega=L/r^2\neq 0$ hold at a point 
on the trajectory where $r=r^*$. 

A {\em turning point} is a finite radial value $r=r_*$ 
at which the effective potential is equal to the energy $E$. 
Hence the set of all turning points is given by the roots (if any) of 
the energy equation
\begin{equation}\label{turningpoint}
0= U_\eff(r_*) -E = \tfrac{1}{2} L^2 r_*^{-2} + U(r_*)  -E - U(r_\eq),
\quad
0\leq r_*<\infty . 
\end{equation}
A solution of the polar \eom/ \eqref{polar-eom} possesses a turning point
if (and only if) $v=0$ holds at a point on the trajectory where $r=r^*$, 
since $E-U_\eff(r)=\tfrac{1}{2}v^2$ from equations \eqref{E} and\eqref{Ueff}. 
Note that a turning point $r=r_*$ will coincide with an inertial point $r=r^*$
whenever the energy has the value $E=U_\eff(r^*)$. 

For all potentials $U(r)$ of physical interest, 
at least one turning point or one inertial point can be assumed to exist 
for every solution of the polar \eom/ \eqref{polar-eom}. 
However, if a solution is circular, 
then the integral expressions \eqref{I3=Theta} and \eqref{I4=T} 
cannot even be defined since $r$ is constant, 
and therefore only the first integrals for angular momentum \eqref{L} and energy \eqref{E} exist. 
Consequently, the question of determining the constants $\Theta_0$ and $T_0$ 
needs to be considered only for non-circular solutions. 

A natural way to determine $\Theta_0$ and $T_0$ is by requiring that 
the respective values of the first integrals \eqref{I3=Theta} and \eqref{I4=T}
for each non-circular solution of the polar \eom/ \eqref{polar-eom} 
are given by the conditions $\Theta|_{r=r_0}=\theta_0$ and $T|_{r=r_0}=t_0$ 
where either $r(t_0)=r_0=r^*$ and $\theta(t_0)=\theta_0$ is an inertial point on the trajectory, 
or $r(t_0)=r_0=r_*$ and $\theta(t_0)=\theta_0$ is a turning point on the trajectory. 
These conditions lead directly to the expressions
\begin{equation}\label{Theta}
\Theta = \theta - L\int^r_{r_0} \frac{\sgn(v)}{\sqrt{ 2(E+U(r_\eq)-U(r))r^4 -L^2 r^2}}\,dr 
\end{equation}
and
\begin{equation}\label{T}
T = t- \int^r_{r_0} \frac{\sgn(v)}{\sqrt{ 2(E+U(r_\eq)-U(r)) -L^2 r^{-2}}}\,dr 
\end{equation}
where
\begin{equation}\label{r0=tp}
r_0=r_*,
\quad
r_*^{2}(U(r_*) -E - U(r_\eq)) = \tfrac{1}{2} L^2
\end{equation}
or 
\begin{equation}\label{r0=ip}
r_0=r^*,
\quad
r^*{}^{3}U'(r^*)  =L^2 . 
\end{equation}
Note that the choice of $r_0$ in these integrals corresponds to 
specifying the values for the constants $\Theta_0$ and $T_0$. 

Thus we have proved the following main result. 

\begin{thm}\label{allI}
For the polar \eom/ \eqref{polar-eom} of general central force dynamics:
\newline 
{\rm (1)}
$L$ and $E$ are well-defined first integrals for all solutions. 
($L$ depends solely on $r,\omega$; 
$E$ depends solely on $v$ and $U_\eff(r)$). 
\newline 
{\rm (2)}
$\Theta$ and $T$ are well-defined first integrals for all non-circular solutions. 
(Both $\Theta$ and $T$ depend solely on $L$, $E$, $\sgn(v)$ and $U_\eff(r)$.)
\newline 
{\rm (3)}
$L$, $E$, $\Theta$ are functionally independent \coms/. 
\newline 
{\rm (4)}
Every first integral is a function of $L$, $E$, $\Theta$, $T$,
and every \com/ is a function of $L$, $E$, $\Theta$.  
\end{thm}

We emphasize that these four first integrals, 
given by equations \eqref{L}, \eqref{E}, \eqref{Theta}, \eqref{T}, 
can be directly verified to obey 
\begin{equation}
\frac{dL}{dt}\big|_\soln =\frac{dE}{dt}\big|_\soln =\frac{d\Theta}{dt}\big|_\soln =\frac{dT}{dt}\big|_\soln =0 . 
\end{equation}

We will see later that the first integral $\Theta$ is directly related to the angle of the \LRL/ vector in the plane of motion for non-circular solutions. 

\subsection{Evaluation using turning points}

Each solution $(r(t),\theta(t))$ of the polar \eom/ \eqref{polar-eom}
describes a trajectory in the plane of motion. 
The shape of a trajectory is a curve $\theta=f(r,L,E,\Theta)$ 
parameterized by the values of the angular momentum $L$ and the energy $E$, 
as well as the angle $\Theta$,
with $r$ belonging to some specified radial domain $r_\min\leq r\leq r_\max$. 

The angular first integral \eqref{Theta} gives an algebraic (quadrature) 
equation for the curve, 
\begin{equation}\label{polarcurve}
\theta = f(r,L,E,\Theta) 
= \Theta + L\int^r_{r_0} \frac{\sgn(v)}{\sqrt{ 2(E+U(r_\eq)-U(r))r^4 -L^2 r^2}}\,dr , 
\end{equation}
while the motion along the curve is implicitly given by 
the temporal first integral \eqref{T}, 
\begin{equation}\label{polarmotion}
t = T +\int^r_{r_0} \frac{\sgn(v)}{\sqrt{ 2(E+U(r_\eq)-U(r)) -L^2 r^{-2}}}\,dr .
\end{equation}

If the angular momentum $L=r^2\omega$ is equal to zero,
then the trajectory is purely radial, 
since $\omega=0$ implies $\theta(t)=\theta(0)$ is constant. 
The equation of the curve \eqref{polarcurve} thereby reduces to 
an unbounded radial line $\theta=\Theta=\const$ through $r=0$, 
where the angular first integral \eqref{Theta} 
evaluated on the radial trajectory $(r(t),\theta(t))$ 
is given by $\Theta=\theta(0)$,
independently of the choice of $r_0$. 
Thus, in this case, 
$\Theta$ is the angle of the radial trajectory in the plane of motion. 

The physically more interesting trajectories are non-radial,
which occur if the angular momentum has a non-zero value, $L=r^2\omega\neq 0$, 
so then $\theta(t)$ is no longer constant. 
There are two different primary types of non-radial trajectories:
bounded and unbounded. 
A {\em bounded trajectory} has a radial domain given by 
$0\leq r_\min\leq r_\max<\infty$,
whereas the radial domain for an {\em unbounded trajectory} is given by
$0\leq r_\min<r_\max=\infty$. 
A crucial feature of both types of trajectories is 
the number of turning points that occur in the radial domain. 
When a trajectory has $r_\min>0$, 
any point with $r=r_\min$ on the trajectory 
corresponds to a turning point given by $r_*=r_\min$. 
Similarly, when a trajectory has $r_\max<\infty$, 
any point with $r=r_\max$ on the trajectory 
corresponds to a turning point given by $r_*=r_\max$. 
Bounded non-radial trajectories are thus characterized 
by a radial domain having precisely two turning points 
$r_{*}=r_\min$ and $r_{*}=r_\max$ when the trajectory is non-circular,  
or a single turning point $r_{*}=r_\min=r_\max$ 
in the special case when the trajectory is circular. 
Unbounded non-radial trajectories are characterized by 
having a radial domain with either only one turning point, $r_{*}=r_\min$, 
or no turning points. 

Any turning point that occurs on a non-circular trajectory $(r(t),\theta(t))$
is a radial extremum which is 
either a {\em periapsis} of the trajectory where $r=r_{*}=r_\min>0$,
or an {\em apoapsis} of the trajectory where $r=r_{*}=r_\max<\infty$. 
These apses may occur at any number of distinct angles
(with the same radial distance $r=r_{*}$) 
depending on the shape of the trajectory. 
When a non-circular trajectory $(r(t),\theta(t))$ 
passes through any apsis, $(r_{*},\theta_{*})$, 
the radial speed $v(t)$ will change its sign. 
From equation \eqref{polarcurve} we then have the following result.

\begin{prop}\label{reflectsymm}
For any solution of the polar \eom/ \eqref{polar-eom} 
yielding a non-circular trajectory, 
the curve \eqref{polarcurve} determined by the shape of the trajectory  
is locally symmetric (under reflection) 
around the radial line connecting the origin $r=0$ to any apsis point 
in the plane of motion. 
\end{prop}

We will refer to the radial line $\theta=\theta_{*}$ 
determined by a given apsis $(r_{*},\theta_{*})$ on a non-circular trajectory 
as an {\em apsis line}. 
Note that every apsis on a given trajectory 
is a turning point of the effective potential, 
but the set of turning points \eqref{turningpoint} 
may include points that do not occur on the trajectory.

It is clear that turning points are important in 
the evaluation of the angular and temporal first integrals \eqref{Theta} and \eqref{T} 
since the expression 
\begin{equation}\label{v}
\sqrt{2(E+U(r_\eq)-U(r)) -L^2 r^{-2}}=|v|
\end{equation}
appearing in these integrals will vanish at all turning points $r=r_{*}$ 
that occur on a given non-circular trajectory $(r(t),\theta(t))$. 

\subsection{Piecewise property (``multiplicities'') and trajectory shapes}
\label{interpret}

There are three different types of non-circular trajectories which may arise,
depending on the number of apis points. 
The physical and mathematical properties of 
the angular and temporal first integrals \eqref{Theta} and \eqref{T} 
differ in each case. 

First, suppose a non-circular trajectory $(r(t),\theta(t))$ possesses no apsis. 
This implies that the radial domain of the trajectory 
contains no turning points and hence the trajectory is unbounded. 
Consequently, 
the equation \eqref{polarcurve} of the curve 
describing the shape of the trajectory 
will be valid on the whole radial domain, 
and similarly the angular first integral \eqref{Theta} 
will give a unique value for $\Theta$ 
when it is evaluated on any part of the trajectory,
regardless of the choice of $r_0$ in the integral.
The most physically meaningful and mathematically simple choice for $r_0$
will be an inertial point, $r_0=r^*$, 
given by either the maximum or the minimum extremum of the effective potential. 
With this choice, 
the value of $\Theta$ will be the angle $\theta=\theta^{*}=\Theta$ of the point 
on the trajectory at which the radial speed $v$ is an extremum,
while the value of $T$ given by the temporal first integral \eqref{T}
will be the time $t=t^{*}=T$ at which this point is reached on the trajectory. 

Next, suppose a non-circular trajectory possesses a single apsis point,
$(r,\theta)=(r_{*},\theta_{*})$. 
The radial domain therefore contains exactly one turning point, $r=r_{*}$,
and the trajectory is thus unbounded. 
In this case the first integrals \eqref{Theta} and \eqref{T} 
are most naturally defined by having $r_0=r_{*}$ chosen to be the turning point
(corresponding to the apsis). 
Then the equation \eqref{polarcurve} of the curve 
which describes the shape of the trajectory 
will be valid on the whole radial domain. 
Because this domain contains a single turning point, $r=r_{*}$, 
the curve will be globally reflection-symmetric 
around the corresponding apsis line $\theta=\theta_{*}$, 
and the resulting two halves of the curve will come from separately putting 
$\sgn(v)=+1$ and $\sgn(v)=-1$ in equation \eqref{polarcurve}.
Moreover, 
the angular first integral \eqref{Theta} will give a unique value
for $\Theta$ when it is evaluated on any part of the trajectory. 
The physical meaning of this value is that it will be the angle 
$\theta=\theta_{*}=\Theta$ at which the apsis occurs on the trajectory 
in the plane of motion. 
The same argument shows that the temporal first integral \eqref{T}
gives a unique value for $T$ which will be the time $t=t_{*}=T$ 
at which the apsis $r=r_{*}$ is reached on the trajectory. 

Finally, suppose a non-circular trajectory $(r(t),\theta(t))$ 
possesses multiple apsis points (apsides). 
The radial domain of the trajectory thereby contains exactly two turning points,
which are the endpoints of the domain. 
Correspondingly, the trajectory is bounded. 
The equation \eqref{polarcurve} of the curve 
then will divide up into separate pieces that are determined 
in the following way by the apis points. 
Since $r(t)$ is increasing nearby any periapsis point, 
the trajectory must pass through an apoapsis point
before reaching another periapsis point. 
Likewise, since $r(t)$ is decreasing nearby any apoapsis point, 
the trajectory must pass through an periapsis point
before reaching another apoapsis point. 
Hence, 
multiple apsis points come in pairs consisting of both a periapsis and an apoapsis. 
Let $(r,\theta)=(r_{*1},\theta_{*1})$ and $(r,\theta)=(r_{*2},\theta_{*2})$ 
be any pair of adjacent periapsis and apoapsis points on the trajectory,
whereby $0<r_\min=r_{*1}<r_{*2}=r_\max<\infty$. 
Consider the piece of the curve starting at the periapsis point, 
as defined by choosing $r_0=r_{*1}$ in the curve equation \eqref{polarcurve}, 
with $\Theta=\theta_{*1}$. 
This piece of the curve has the radial domain $r_{*1}\leq r\leq r_{*2}$,
where the other endpoint $r=r_{*2}$ is an apoapsis point. 
Since the curve is locally reflection-symmetric 
around the apoapsis line $\theta=\theta_{*1}$ in the plane of motion,
the next piece of the curve is defined by choosing $r_0=r_{*2}$ in equation \eqref{polarcurve}, 
with $\Theta=\theta_{*2}$,
where the radial domain is $r_{*2}\geq r\geq r_{*1}$. 
Note that $\sgn(v)$ differs on these two pieces of the curve. 
Clearly, this process can be continued to piece together the entire curve,
which yields the complete trajectory $(r(t),\theta(t))$. 

Thus, for any bounded non-circular trajectory with multiple apsis points, 
it follows that the angular first integral \eqref{Theta} will be multi-valued
when it is evaluated on different parts of the trajectory, 
corresponding to the piecewise composition of the curve \eqref{polarcurve}. 
The same property will hold for the temporal first integral \eqref{T},
since it is uses the same value of $r_0$ that is chosen 
in the angular the first integral \eqref{Theta}. 
Each pair of values $\Theta$ and $T$ given by these first integrals 
are the angle and the time at which the trajectory reaches each apsis. 

As a consequence of \propref{reflectsymm}, 
the angular separation between two successive apsis points on
a bounded non-circular trajectory is the same on all parts of the trajectory. 
From the curve equation \eqref{polarcurve}, 
it is straightforward to get an integral expression for this angular separation.

\begin{prop}\label{apsisangle}
For any solution of the polar \eom/ \eqref{polar-eom} 
yielding a bounded non-circular trajectory, 
the angular separation between any two successive 
periapsis points or apoapsis points on the trajectory is given by 
\begin{equation}\label{precessangle}
\Delta\theta = 
2L\int^{r_\max}_{r_\min} \frac{1}{\sqrt{ 2(E+U(r_\eq)-U(r))r^4 -L^2 r^2}}\,dr 
\quad(\text{modulo $2\pi$}),
\end{equation}
and the corresponding time interval is given by 
\begin{equation}\label{precesstime}
\Delta t = 
2\int^{r_\max}_{r_\min} \frac{\sgn(v)}{\sqrt{ 2(E+U(r_\eq)-U(r)) -L^2 r^{-2}}}\,dr . 
\end{equation}
These expressions are related to the first integrals $\Theta$ and $T$ by 
\begin{equation}\label{separations}
\tfrac{1}{2}\Delta\theta = \sgn(v)(\Theta|_{r_0=r_\max}-\Theta|_{r_0=r_\min}), 
\quad
\tfrac{1}{2}\Delta t = \sgn(v)(T|_{r_0=r_\max}-T|_{r_0=r_\min}).
\end{equation}
\end{prop}

If the angular separation $\Delta\theta$ is a rational multiple of $2\pi$, 
then the angular first integral \eqref{Theta} 
will yield a finite number of distinct values $\Theta$ modulo $2\pi$. 
In this case, the curve describing the shape of the trajectory is closed.
Note that the apsis points of the trajectory in the plane of motion 
are precessing unless $\Delta\theta$ is exactly equal to $2\pi$. 

In contrast, 
if the angular separation $\Delta\theta$ is an irrational multiple of $2\pi$, 
then the angular first integral \eqref{Theta} 
will yield an infinite number of distinct values $\Theta$ modulo $2\pi$. 
This means that the curve describing the shape of the trajectory is open, 
such that the apsis points of the trajectory are precessing 
in the plane of motion. 

For both kinds of (bounded non-circular) trajectories, 
the temporal first integral \eqref{T} yields a periodic infinite sequence of values $T$.

\section{Examples}
\label{examples}

The results in \thmref{allI}, \propref{reflectsymm} and \propref{apsisangle}, 
will now be illustrated for two examples of central forces. 

\subsection{Inverse-square force}
\label{inversesquare}
Consider the central force
\begin{equation}\label{invsqF}
F=-kr^{-2}
\end{equation}
with the potential 
\begin{equation}
U(r) = -kr^{-1}
\end{equation}
This force will be attractive if $k>0$, \eg/ planetary motion, 
or repulsive if $k<0$, \eg/ Coulomb scattering of charged particles. 
In either case, 
the potential has only one equilibrium point, 
$r_{\eq} =\infty$. 
Since $U(r_{\eq})=0$, the effective potential is given by 
\begin{equation}\label{kepler-effU}
U_\eff(r) = \tfrac{1}{2} L^2 r^{-2} -kr^{-1},
\quad
L\neq 0 .
\end{equation}
Thus 
\begin{equation}
E = \tfrac{1}{2}v^2 +\tfrac{1}{2}L^2r^{-2}-kr^{-1}
\end{equation}
is the energy first integral \eqref{E}. 
All trajectories can be classified qualitatively from the equation $E=U_\eff(r)$.

In the repulsive case $k<0$, $U_\eff(r)$ has no extremum, 
and $E$ is non-negative. 
Hence all trajectories are unbounded. 

In the attractive case $k>0$, $U_\eff(r)$ has one extremum,
which is a negative minimum, $U_\eff^\min= -\tfrac{1}{2}(k/L)^2$. 
Hence, for $L\neq0$,  
all trajectories with $E\geq 0$ are unbounded,
while all trajectories with $0>E\geq U_\eff^\min$ are bounded, 
and bounded trajectories with $E=U_\eff^\min$ are circular. 

We now evaluate the angular and temporal first integrals \eqref{Theta} and \eqref{T} for the case $k>0$, $L\neq0$, \ie/ the Kepler problem. 
We will need the relation 
\begin{equation}\label{kepler-v}
|v|= \sqrt{2E-L^2r^{-2}+2kr^{-1}} . 
\end{equation}

\subsubsection{Turning points for the Kepler problem}
The turning points $r=r_*$ of the effective potential \eqref{kepler-effU}
are obtained from the energy equation \eqref{turningpoint}. 
This is a quadratic equation
\begin{equation}
0=2Er^2+2kr-L^2
\end{equation}
with the discriminant $D= 4(k^2-2EL^2)$. 
Turning points exist only when $D\geq 0$,
where $D=0$ determines the minimum energy
\begin{equation}
E_\min = -\frac{k^2}{2L^2} <0 . 
\end{equation}
Since the trajectories with $E=E_\min$ are circular, 
only the trajectories with $E>E_\min$ need to be considered
(as the first integrals $\Theta$ and $T$ exist only for non-circular trajectories). 

For $0>E>E_\min$, the turning points are given by 
\begin{equation}\label{kepler-negE-tp}
r_{*\pm} = \frac{k\pm\sqrt{k^2-2|E|L^2}}{2|E|} 
\end{equation}
The angular first integral \eqref{Theta} with $r_0=r_{*\pm}$ 
can be straightforwardly evaluated to yield
\begin{equation}\label{kepler-negE-Theta}
\Theta_\pm
=\theta+\arctan\left(\frac{L^2-kr}{Lvr}\right) \pm\sgn(vL)\frac{\pi}{2}
\end{equation}
after use of the relation \eqref{kepler-v}.
Similarly, 
the temporal first integral \eqref{T} with $r_0=r_{*\pm}$ yields
\begin{equation}\label{kepler-negE-T}
T_\pm =t-\frac{rv}{2|E|}+\frac{k}{(2|E|)^{3/2}}\Big( \arctan\left(\frac{2|E|r-k}{\sqrt{2|E|}vr}\right)\mp \sgn(v) \frac{\pi}{2} \Big) . 
\end{equation}
From the first integral \eqref{kepler-negE-Theta}, 
trajectories with $0>E>E_\min$ are algebraically described by curves 
\begin{equation}\label{kepler-negE-shape}
r(\theta)=\frac{L^2}{k\mp\sqrt{k^2-2|E|L^2}\cos(\Theta_\pm-\theta)}
\end{equation}
on the angular domain 
\begin{equation}\label{kepler-negE-domain}
-\tfrac{1}{2}\Delta\theta +\Theta_\pm \leq\theta \leq \tfrac{1}{2}\Delta\theta + \Theta_\pm 
\end{equation}
where, as shown by \propref{apsisangle}, 
\begin{equation}
\tfrac{1}{2}\Delta\theta = \sgn(v)(\Theta_+ -\Theta_-) = \pi 
\end{equation}
is the angular separation between the two turning points on the curve.
These curves are ellipses \cite{GolPooSaf},
which are bounded and have $r_\max= r_{*+}$ and $r_\min =r_{*-}$, 
such that $r=0$ is one focus of the ellipse in the plane of motion. 
Moreover, in accordance with \propref{reflectsymm}, 
each ellipse is reflection-symmetric around a radial line 
connecting the origin $r=0$ to either of the points $r=r_{*\pm}$.  
The resulting two pieces of the ellipse are 
each defined on the radial domain $r_\min\leq r\leq r_\max$
and differ by $\sgn(v)\gtrless 0$
(where $\sgn(v)=0$ at each turning point). 

On each elliptic trajectory, 
both first integrals $\Theta_\pm$ and $T_\pm$ are piecewise functions. 
Their physical and geometrical meaning depends on the choice of 
the radial value $r_0=r_+$ or $r_0=r_-$. 
If $r_0=r_+$ then $(r(T_+),\theta(T_+))=(r_\max,\Theta_+)$ is an apoapsis point 
on the trajectory, 
or instead if $r_0=r_-$ then $(r(T_-),\theta(T_-))=(r_\min,\Theta_-)$ is a periapsis point.
In both cases, 
$\Theta_\pm$ and $T_\pm$ are continuous at the apsis point $r=r_{*\pm}$, 
but have a jump discontinuity at the opposite apsis $r=r_{*\mp}$. 
The jump in $\Theta_{\pm}$ is equal to 
\begin{equation}\label{kepler-negE-precessangle}
\Delta\theta = 2\pi
\end{equation}
which is the angular separation \eqref{precessangle}
between two successive periapsis points 
or two successive apoapsis points on the trajectory. 
Since $\Delta\theta$ is exactly $2\pi$, 
these trajectories are not precessing,
and hence both of the angular first integrals $\Theta_\pm$ are single-valued modulo $2\pi$.  
The jump in $T_\pm$ is simply the period of a single closed orbit, 
\begin{equation}\label{kepler-negE-precesstime}
\Delta t = \frac{\pi k}{\sqrt{2|E|^3}} . 
\end{equation}
Both of the temporal first integrals $T_\pm$ thus yield a sequence of times 
such that the time interval \eqref{precesstime} is the period $\Delta t$. 

Therefore, 
when the angular first integral $\Theta_-$ and temporal first integral $T_-$
are evaluated at any point $(r(t),\theta(t))$ on an elliptic trajectory \eqref{kepler-negE-shape}, 
we obtain the angle of the radial line that intersects the periapsis point 
and the time at which this point is reached on the trajectory. 
Likewise the angular first integral $\Theta_+$ and temporal first integral $T_+$
evaluated at any point $(r(t),\theta(t))$
yield the angle of the radial line that intersects apoapsis point 
and the time at which this point is reached on the elliptic trajectory. 

Next, for $E=0$, there is just a single turning point
\begin{equation}\label{kepler-E=0-tp}
r_*=\frac{L^2}{2k} . 
\end{equation}
The angular first integral \eqref{Theta} with $r_0=r_*$ 
is straightforward to evaluate, 
\begin{equation}\label{kepler-E=0-Theta}
\Theta=\theta+\arctan\left(\frac{L^2-kr}{Lvr}\right)-\sgn(vL)\frac{\pi}{2} . 
\end{equation}
Similarly, evaluation of the temporal first integral \eqref{T} with $r_0=r_*$ 
yields
\begin{equation}\label{kepler-E=0-T}
T = t-\frac{rv(L^2-kr)}{3k^2} . 
\end{equation}
From the first integral \eqref{kepler-E=0-Theta}, 
trajectories with $E=0$ are algebraically described by curves 
\begin{equation}\label{kepler-E=0-shape}
r(\theta)=\frac{L^2}{k(1+\cos(\Theta-\theta))}
\end{equation}
on the angular domain 
\begin{equation}\label{kepler-E=0-domain}
-\pi +\Theta \leq\theta \leq \pi + \Theta . 
\end{equation}
These curves \eqref{kepler-E=0-shape} are parabolas \cite{GolPooSaf} 
with the focus at $r=0$, such that $r_\min=r_*$. 
In accordance with \propref{reflectsymm}, 
each parabola is reflection-symmetric around the radial line 
connecting the origin $r=0$ to the periapsis point $r=r_*$ on the trajectory. 
The resulting two pieces of the parabola are 
each defined on the radial domain $r_\min\leq r<\infty$
and differ by $\sgn(v)\gtrless 0$
(where $\sgn(v)=0$ at the turning point). 

The first integrals $\Theta$ and $T$ are piecewise functions on each parabola,
such that $(r(T),\theta(T))=(r_\min,\Theta)$ is the periapsis point. 
Both $\Theta$ and $T$ are continuous at this point. 
Thus, when these first integrals are evaluated at any point $(r(t),\theta(t))$ 
on a parabolic trajectory \eqref{kepler-E=0-shape}, 
we obtain the angle of the radial line that intersects the periapsis point 
and the time at which this point is reached on the trajectory. 

Last, for $E>0$, there is again a single turning point
\begin{equation}\label{kepler-posE-tp}
r_*=\frac{\sqrt{k^2+2EL^2}-k}{2E} . 
\end{equation}
Evaluation of the angular first integral \eqref{Theta} with $r_0=r_*$ 
straightforwardly yields the same expression \eqref{kepler-E=0-Theta}
as in the case $E=0$. 
The temporal first integral \eqref{T} with $r_0=r_*$ gives
\begin{equation}\label{kepler-posE-T}
T =t-\frac{1}{2E}\Big( rv+\frac{k}{\sqrt{2E}}\arctanh\left(\frac{\sqrt{2E}vr}{2Er+k}\right) \Big) . 
\end{equation}
From the first integral \eqref{kepler-E=0-Theta} combined with the relation \eqref{kepler-v}, 
trajectories with $E>0$ are algebraically described by curves 
\begin{equation}\label{kepler-posE-shape}
r(\theta) = \frac{L^2}{k+\sqrt{k^2+2EL^2}\cos(\Theta-\theta)}
\end{equation}
on the angular domain 
\begin{equation}\label{kepler-posE-domain}
- \frac{\pi}{2} -\arctan\left(\sqrt{2E}|L|/k\right) +\Theta 
\leq\theta \leq 
\frac{\pi}{2} +\arctan\left(\sqrt{2E}|L|/k\right) +\Theta . 
\end{equation}
These trajectories are hyperbolas \cite{GolPooSaf},
which are unbounded and have $r=0$ as the focus.
The properties of the angular first integral $\Theta$ and temporal first integral $T$
for a hyperbolic trajectory are the same as in the parabolic case. 

\subsubsection{Inertial points for the Kepler problem}
The inertial points $r=r^*$ of the effective potential \eqref{kepler-effU}
are obtained from the effective force equation \eqref{inertialpoint}. 
This reduces to a linear equation
\begin{equation}
0=L^2-kr
\end{equation}
giving 
\begin{equation}\label{kepler-ip}
r^*=\frac{L^2}{k} . 
\end{equation}
From \propref{reflectsymm}, 
every trajectory with $L\neq 0$ will contain two inertial points $r=r^*$,
differing by $\sgn(v)\gtrless 0$. 
In particular, the radial speed at the these points is given by 
\begin{equation}
v^*=\frac{\pm\sqrt{k^2-2|E|L^2}}{|L|}
\end{equation}
by combining equations \eqref{kepler-ip} and \eqref{kepler-v}.

For $0>E>E_\min$, 
the angular first integral \eqref{Theta} with $r_0=r^*$ 
can be straightforwardly evaluated to yield
\begin{equation}\label{kepler-ip-negE-Theta}
\Theta=\theta + \arctan\left(\frac{L^2-kr}{Lvr}\right) . 
\end{equation}
Similarly, the temporal first integral \eqref{T} with $r_0=r^*$ gives
\begin{equation}\label{kepler-ip-negE-T}
\begin{aligned}
T = &
t +\frac{rv}{2|E|} - \frac{\sgn(v)|L|\sqrt{k^2-2|E|L^2}}{2|E|k} 
\\&\qquad
+\frac{k}{(2|E|)^{3/2}} \Big( \arctan\left(\frac{k-2|E|r}{\sqrt{2|E|}vr}\right)
-\sgn(v)\arctan\left(\frac{\sqrt{k^2-2|E|L^2}}{\sqrt{2|E|}L}\right) \Big) . 
\end{aligned}
\end{equation}
From the first integral \eqref{kepler-ip-negE-Theta}, 
the curves describing trajectories with $E>0$ are algebraically given by 
\begin{equation}\label{kepler-ip-negE-shape}
r(\theta) = \frac{L^2}{k+\sqrt{k^2-2|E|L^2}\sin(\Theta-\theta)} . 
\end{equation}
These curves are ellipses \eqref{kepler-negE-shape}, 
with a different angular parameterization compared to the turning point case. 

On each elliptic trajectory, 
both first integrals $\Theta$ and $T$ are piecewise functions 
which are continuous at the two inertial points $r=r^*$ 
but have a jump discontinuity at the two apsis points $r=r_\min$ and $r=r_\max$. 
The jump in $\Theta$ is equal to 
\begin{equation}\label{kepler-ip-negE-precessangle}
\Delta\Theta= \tfrac{1}{2}\Delta\theta = \pi 
\quad(\text{modulo $2\pi$}) 
\end{equation}
which is the angular separation between the two apsis points. 
Similarly, the jump in $T$ is half of the period of a single closed orbit, 
\begin{equation}\label{kepler-ip-negE-precesstime}
\Delta T= \tfrac{1}{2}\Delta t = \frac{\pi k}{\sqrt{4|E|^3}} . 
\end{equation}
The physical and geometrical meaning of these jumps is directly related to 
the division of an elliptic trajectory into two reflection-symmetric pieces
around the radial line (semi-major axis) through the two apsis points. 
Note that the two inertial points $r=r^*$ lie on a perpendicular line 
(semi-minor axis) with respect to the apsis points. 
Therefore, when the first integrals $\Theta$ and $T$ are evaluated 
at any point $(r(t),\theta(t))$ on the piece with $\sgn(v)\gtrless 0$, 
we obtain $(r(T),\theta(T))=(r^*,\Theta)$ 
which yields the angle of the radial line that intersects 
the inertial point on that piece of the trajectory, 
and the time at which this point is reached. 

Next, for $E=0$, 
the angular first integral \eqref{Theta} with $r_0=r^*$ is again given by 
the expression \eqref{kepler-ip-negE-Theta},
while the temporal first integral \eqref{T} with $r_0=r^*$ yields
\begin{equation}\label{kepler-ip-E=0-T}
T=t-\sgn(v)\left(\frac{r|v|(L^2+kr)-2|L|^3}{3k^2}\right) . 
\end{equation}
The resulting curves that describe trajectories with $E=0$ 
are algebraically given by 
\begin{equation}\label{kepler-ip-E=0-shape}
r(\theta) = \frac{L^2}{k(1+\sin(\Theta-\theta))} . 
\end{equation}
These curves are parabolas \eqref{kepler-negE-shape}, 
with an angular parameterization that is shifted by $\pi/2$ 
compared to the turning point case. 
The two inertial points on each parabola are related by reflection symmetry 
through the radial line that connects the periapsis point $r=r_\min$ to the origin $r=0$. 

On each parabolic trajectory, 
both first integrals $\Theta$ and $T$ are piecewise functions 
which are continuous at the two inertial points $r=r^*$ 
but have a jump discontinuity at the periapsis point $r=r_\min$,
corresponding to the division of the trajectory into two reflection-symmetric pieces. 
The jumps in $\Theta$ and $T$ are given by the same respective values 
\eqref{kepler-ip-negE-precessangle} and \eqref{kepler-ip-negE-precesstime}
as in the elliptic case. 
Therefore, when the first integrals $\Theta$ and $T$ are evaluated 
at any point $(r(t),\theta(t))$ on the piece of a parabolic trajectory with $\sgn(v)\gtrless 0$, 
we obtain $(r(T),\theta(T))=(r^*,\Theta)$ 
which yields the angle of the radial line that intersects 
the inertial point on that piece of the trajectory, 
and the time at which this point is reached. 

Last, for $E>0$, 
the same expression \eqref{kepler-ip-negE-Theta} is obtained
for the angular first integral \eqref{Theta} with $r_0=r^*$, 
while the expression for the temporal first integral \eqref{T} with $r_0=r^*$ 
is given by 
\begin{equation}\label{kepler-ip-posE-T}
\begin{aligned}
T =&
t + \frac{rv}{2E} -\frac{\sgn(v)|L|\sqrt{k^2+2EL^2}}{2kE} 
\\&\qquad
+\frac{k}{(2E)^{3/2}} \Big( 
\arctanh\left(\frac{\sqrt{2E}vr}{2Er+k}\right) - \sgn(v)\arctanh\left(\frac{\sqrt{2E}|L|}{\sqrt{k^2+2EL^2}}\right) \Big) . 
\end{aligned}
\end{equation}
Trajectories with $E>0$ are algebraically described by the curves
\begin{equation}\label{kepler-ip-posE-shape}
r(\theta) = \frac{L^2}{k+\sqrt{k^2+2EL^2}\sin(\Theta-\theta)}
\end{equation}
which are hyperbolas \eqref{kepler-posE-shape}, 
with an angular parameterization that is shifted by $\pi/2$ 
compared to the turning point case. 
The two inertial points on each hyperbola are related by reflection symmetry 
through the radial line that connects the periapsis point $r=r_\min$ to the origin $r=0$. 

The physical and geometrical meaning of the first integrals $\Theta$ and $T$ 
is similar to the parabolic case. 
Evaluation of $\Theta$ and $T$ 
at any point $(r(t),\theta(t))$ on the piece of a hyperbolic trajectory with $\sgn(v)\gtrless 0$
yields the angle of the radial line that intersects 
the inertial point on that piece of the trajectory, 
and the time at which this point is reached. 
At the periapsis point $r=r^*$, 
$\Theta$ jumps by the value \eqref{kepler-ip-negE-precessangle},
while the jump in $T$ is given by 
\begin{equation}\label{kepler-ip-posE-precesstime}
\Delta T = \frac{k}{\sqrt{2E^3}} \arctanh\left(\frac{\sqrt{2E}|L|}{\sqrt{k^2+2EL^2}}\right) . 
\end{equation}

\subsection{Inverse square force with cubic corrections}
Consider an inverse-cube correction to the inverse-square central force \eqref{invsqF}, 
\begin{equation}\label{invsqcorrF}
F(r) = -kr^{-2} -\kappa r^{-3}
\end{equation}
which has the potential 
\begin{equation}
U(r) = -\frac{k}{r}-\frac{\kappa}{2r^2} . 
\end{equation}
Both terms in this force will be attractive if $k>0$ and $\kappa>0$, 
in which case the potential has only one equilibrium point, 
$r_{\eq} =\infty$. 
Since $U(r_{\eq})=0$, the effective potential is given by 
\begin{equation}\label{newton-effU}
U_{\eff}(r) = \frac{L^2-\kappa}{2r^2} -\frac{k}{r} . 
\end{equation}
Then 
\begin{equation}
E = \tfrac{1}{2}v^2+\tfrac{1}{2}(L^2-\kappa)r^{-2}-kr^{-1}
\end{equation}
is the energy first integral \eqref{E}. 
All trajectories can be classified qualitatively from the equation $E=U_\eff(r)$.

If $\kappa \geq L^2$, the correction term will dominate the angular momentum term. 
In this case, the effective potential will have no extrema,
and consequently there are no bounded trajectories,
while all unbounded trajectories pass through the origin $r=0$. 

Thus, hereafter we will take 
\begin{equation}
0<\kappa < L^2 . 
\end{equation}
In this case, 
the properties of the effective potential are similar to the Kepler case. 
There is one extremum, which is a negative minimum, 
$U_\eff^\min= -\tfrac{1}{2}k^2/(L^2-\kappa)^2$.
As a consequence, 
all trajectories with $E\geq 0$ are unbounded,
while all trajectories with $0>E\geq U_\eff^\min$ are bounded, 
and bounded trajectories with $E=U_\eff^\min$ are circular. 
The main difference compared to Kepler case will be that here 
the unbounded trajectories are open and precessing. 
This is called the Newtonian revolving orbit problem. 

We now evaluate the angular and temporal first integrals \eqref{Theta} and \eqref{T}. 
Note these first integrals exist only for non-circular trajectories. 
We will need the relation 
\begin{equation}\label{newton-v}
|v|= \sqrt{2E +(\kappa-L^2)r^{-2}+2kr^{-1}} . 
\end{equation}

\subsubsection{Turning points for the Newtonian revolving orbit problem}
The turning points $r=r_*$ of the effective potential \eqref{newton-effU}
are given by replacing $L^2$ with $L^2-\kappa$ in the expressions
for the turning points in the Kepler case. 
Likewise, the minimum energy becomes 
\begin{equation}
E_\min = -\frac{k^2}{2(L^2-\kappa)} <0 . 
\end{equation}
Only the trajectories with $E>E_\min$ need to be considered,
since the trajectories with $E=E_\min$ are circular 
(in which case the first integrals $\Theta$ and $T$ do not exist). 

For $0>E>E_\min$, there are two turning points 
\begin{equation}\label{newton-negE-tp}
r_{*\pm} = \frac{k\pm \sqrt{k^2-2|E|(L^2-\kappa)}}{2|E|} . 
\end{equation}
The angular first integral \eqref{Theta} with $r_0=r_{*\pm}$ 
can be straightforwardly evaluated to yield
\begin{equation}\label{newton-negE-Theta}
\Theta_\pm = \theta +\frac{L}{\sqrt{L^2-\kappa}} \arctan\left(\frac{L^2-\kappa -kr}{rv\sqrt{L^2-\kappa}}\right) 
\pm\frac{\sgn(v)L}{\sqrt{L^2-\kappa}}\frac{\pi}{2}
\end{equation}
after use of the relation \eqref{newton-v}.
Similarly, 
the temporal first integral \eqref{T} with $r_0=r_{*\pm}$
yields
\begin{equation}\label{newton-negE-T}
T_\pm = t-\frac{rv}{2|E|} +\frac{k}{(2|E|)^{3/2}}\Big( \arctan\left(\frac{\sqrt{2|E|}r-k}{\sqrt{2|E|}rv}\right) \mp \frac{\pi}{2}\sgn(v) \Big)  . 
\end{equation}
From the first integral \eqref{newton-negE-Theta}, 
trajectories with $0>E>E_\min$ are algebraically described by curves 
\begin{equation}\label{newton-negE-shape}
r(\theta) = \frac{L^2-\kappa}{k \mp \sqrt{k^2-2|E|(L^2-\kappa)}\cos(\sqrt{1-\kappa/L^2}\,(\Theta_\pm-\theta))}
\end{equation}
which are bounded and have $r_\max= r_{*+}$ and $r_\min =r_{*-}$.  
The angular domain of these curves is given by 
\begin{equation}\label{newton-negE-domain}
-\tfrac{1}{2}\Delta\theta +\Theta_\pm \leq\theta \leq \tfrac{1}{2}\Delta\theta + \Theta_\pm 
\end{equation}
where, as shown by \propref{apsisangle}, 
\begin{equation}
\tfrac{1}{2}\Delta\theta = \sgn(v)(\Theta_+ -\Theta_-) 
= \frac{\pi L}{\sqrt{L^2-\kappa}}>\pi
\end{equation}
is the angular separation between the two turning points. 
Since $\Delta\theta$ is greater than $2\pi$,
the curves are composed of pieces, 
such that the angular separation \eqref{precessangle}
between two successive periapsis points $r=r_{*-}$ 
or two successive apoapsis points $r=r_{*-}$ on adjacent pieces is given by 
\begin{equation}\label{newton-negE-precessangle}
\Delta\theta = \frac{2\pi L}{\sqrt{L^2-\kappa}} . 
\end{equation}
Moreover, by \propref{reflectsymm}, adjacent pieces are reflection-symmetric 
around a radial line connecting the origin $r=0$ to either of the points $r=r_{*\pm}$ where the pieces join. 
If $\Delta\theta/(2\pi)$ is a rational number, 
then the curve is closed, consisting of a finite number of pieces. 
If instead $\Delta\theta/(2\pi)$ is an irrational number, 
then the curve is open, consisting of an infinite number of pieces. 
In either case, the curve describes a precessing bounded trajectory. 

On each bounded trajectory, 
both first integrals $\Theta_\pm$ and $T_\pm$ are piecewise functions. 
Their physical and geometrical meaning depends on the choice of 
the radial value $r_0=r_+$ or $r_0=r_-$. 
If $r_0=r_+$ then $(r(T_+),\theta(T_+))=(r_\max,\Theta_+)$ is an apoapsis point 
on the trajectory, 
or instead if $r_0=r_-$ then $(r(T_-),\theta(T_-))=(r_\min,\Theta_-)$ is a periapsis point.
In both cases, 
$\Theta_\pm$ and $T_\pm$ are continuous at the apsis point $r=r_{*\pm}$, 
but have a jump discontinuity at the opposite apsis $r=r_{*\mp}$ \cite{Per}. 
The jump in $\Theta_{\pm}$ is equal to the angular separation \eqref{newton-negE-precessangle} 
between two successive periapsis points 
or two successive apoapsis points on the trajectory. 
The jump in $T_\pm$ is the corresponding time interval between these points, 
\begin{equation}\label{newton-negE-precesstime}
\Delta t = \frac{\pi k}{\sqrt{2|E|^3}}
\end{equation}
which represents the period of a single open orbit. 
Both of the temporal first integrals $T_\pm$ thus yield a sequence of times 
such that the time interval \eqref{precesstime} is the period $\Delta t$. 

Therefore, 
when the angular first integral $\Theta_-$ and temporal first integral $T_-$
are evaluated at any point $(r(t),\theta(t))$ on a piece of the bounded trajectory \eqref{newton-negE-shape}, 
we obtain the angle of the radial line that intersects the periapsis point 
on that piece, and the time at which this point is reached. 
Likewise the angular first integral $\Theta_+$ and temporal first integral $T_+$
evaluated at any point $(r(t),\theta(t))$ on a piece of the bounded trajectory
yield the angle of the radial line that intersects the apoapsis point on that piece,
and the time at which this point is reached. 

Next, for $E=0$, there is a single turning point, 
\begin{equation}\label{newton-E=0-tp}
r_*=\frac{L^2-\kappa}{2k} . 
\end{equation}
The angular first integral \eqref{Theta} with $r_0=r_*$ 
is straightforward to evaluate, 
\begin{equation}\label{newton-E=0-Theta}
\Theta = \theta +\frac{L}{\sqrt{L^2-\kappa}} \arctan\left(\frac{L^2-\kappa -kr}{rv\sqrt{L^2-\kappa}}\right) 
-\frac{\sgn(v)L}{\sqrt{L^2-\kappa}}\frac{\pi}{2} . 
\end{equation}
Similarly, evaluation of the temporal first integral \eqref{T} with $r_0=r_*$ 
yields
\begin{equation}\label{newton-E=0-T}
T = t-\frac{rv(L^2-\kappa-kr)}{3k^2} . 
\end{equation}
From the first integral \eqref{newton-E=0-Theta}, 
trajectories with $E=0$ are algebraically described by curves 
\begin{equation}\label{newton-E=0-shape}
r(\theta) = \frac{L^2-\kappa}{k(1+\cos(\sqrt{1-\kappa/L^2}\,(\Theta_\pm-\theta)))}
\end{equation}
on the angular domain 
\begin{equation}\label{newton-E=0-domain}
-\frac{\pi L}{\sqrt{L^2-\kappa}}
+\Theta \leq\theta \leq 
+ \Theta + \frac{\pi L}{\sqrt{L^2-\kappa}} . 
\end{equation}
These curves \eqref{newton-E=0-shape} are unbounded and have $r_\min=r_*$. 
By \propref{reflectsymm}, 
each curve is reflection-symmetric around the radial line 
connecting the origin $r=0$ to the periapsis point $r=r_*$,
where the two pieces differ by $\sgn(v)\gtrless 0$
(with $\sgn(v)=0$ at the turning point). 

On each unbounded trajectory, 
the first integrals $\Theta$ and $T$ are piecewise functions 
such that $(r(T),\theta(T))=(r_\min,\Theta)$ is the periapsis point. 
Both $\Theta$ and $T$ are continuous at this point. 
Thus, when these first integrals are evaluated at any point $(r(t),\theta(t))$ 
on the trajectory, 
we obtain the angle of the radial line that intersects the periapsis point 
and the time at which this point is reached on the trajectory. 
This is qualitatively the same as the Kepler case. 

Last, for $E>0$, there is again a single turning point, 
\begin{equation}\label{newton-posE-tp}
r_*=\frac{\sqrt{k^2+2E(L^2-\kappa)}-k}{2E} . 
\end{equation}
Evaluation of the angular first integral \eqref{Theta} with $r_0=r_*$ 
straightforwardly yields the same expression \eqref{newton-E=0-Theta}
as in the case $E=0$. 
The temporal first integral \eqref{T} with $r_0=r_*$ gives
\begin{equation}\label{newton-posE-T}
T =t-\frac{1}{2E}\Big( rv+\frac{k}{\sqrt{2E}}\arctanh\left(\frac{\sqrt{2E}vr}{2Er+k}\right) \Big) . 
\end{equation}
From the first integral \eqref{newton-E=0-Theta} combined with the relation \eqref{kepler-v}, 
trajectories with $E>0$ are algebraically described by curves 
\begin{equation}\label{newton-posE-shape}
r(\theta) = \frac{L^2-\kappa}{k + \sqrt{k^2-2|E|(L^2-\kappa)}\cos(\sqrt{1-\kappa/L^2}\,(\Theta_\pm-\theta))}
\end{equation}
on the angular domain 
\begin{equation}\label{newton-posE-domain}
- \frac{\pi}{2} -\arctan\left(\sqrt{2E}\sqrt{L^2-\kappa}|/k\right) +\Theta 
\leq\theta \leq 
\frac{\pi}{2} +\arctan\left(\sqrt{2E}\sqrt{L^2-\kappa}/k\right) +\Theta . 
\end{equation}
The properties of these curves as well as 
the angular first integral $\Theta$ and temporal first integral $T$
are the same as in the case $E=0$. 

\subsubsection{Inertial points for the Newtonian revolving orbit problem}
The effective potential \eqref{newton-effU} has a single inertial point $r=r^*$,
which is given by replacing $L^2$ with $L^2-\kappa$ in the expression \eqref{kepler-ip}
for the inertial point in the Kepler case, 
\begin{equation}\label{newton-ip}
r^*=\frac{L^2-\kappa}{k} . 
\end{equation}
All of the earlier discussion of the first integrals $\Theta$ and $T$ 
using $r_0=r^*$ in the Kepler case 
carries over here in the same way as the discussion using turning points. 
In particular, 
the main qualitative difference compared to the Kepler case is that 
for $0>E>E_\min$ the trajectories are composed of more than two pieces, 
so consequently $\Theta_\pm$ is multi-valued even when the trajectories are closed.

\section{Symmetry formulation}
\label{noether}

The general form of Noether's theorem \cite{BluAnc,Olv} provides 
a one-to-one explicit correspondence between first integrals and dynamical symmetries of a Lagrangian. 
This correspondence arises from two variational identities as follows. 

We start with the polar Lagrangian \eqref{polar-Lagr} 
for the central force \eom/ \eqref{polar-eom},
using the polar variables $t$, $r$, $\theta$, $v$, $\omega$. 
Now consider any vector field \eqref{polar-X}, 
with the characteristic form \eqref{polar-Xhat},
given by components 
$\tau(t,r,\theta,v,\omega)$, $\xi(t,r,\theta,v,\omega)$, and $\psi(t,r,\theta,v,\omega)$. 
The action of this vector field on the Lagrangian 
can be expressed in terms of the \eom/, 
through the variational derivatives \eqref{polar-Lagr-varders} 
after integration by parts, yielding the identity 
\begin{equation}\label{polar-L-id}
\pr\hat\X(\L) = 
P^r \frac{\delta\L}{\delta r}
+ P^\theta \frac{\delta\L}{\delta \theta} 
+ \frac{dS}{dt}, 
\quad
S= P^r \L_v +P^\theta \L_\omega . 
\end{equation}
From this identity, 
a necessary and sufficient condition for the Lagrangian to be invariant 
modulo a total time derivative 
is that the components $P^r$ and $P^\theta$ of the vector field 
have to satisfy the condition 
\begin{equation}\label{polar-L-inv}
P^r \frac{\delta\L}{\delta r}
+ P^\theta \frac{\delta\L}{\delta \theta} 
= \frac{dR}{dt}
\end{equation}
for some function $R(t,r,\theta,v,\omega)$. 
A vector field \eqref{polar-Xhat} having this property is called 
a {\em variational symmetry}. 
Because any total time derivative is annihilated by a variational derivative, 
the extremals of the Lagrangian are preserved by variational symmetries,
and hence every variational symmetry is an infinitesimal symmetry of the \eom/.
Thus, variational symmetries have the characterization 
as infinitesimal symmetries of the \eom/ \eqref{polar-eom} 
such that the invariance condition 
\begin{equation}\label{polar-varX}
\pr\hat\X(\L) = \frac{d(S+R)}{dt},
\quad
S= P^r \L_v +P^\theta \L_\omega
\end{equation}
holds for some function $R(t,r,\theta,v,\omega)$. 
Note that a variational symmetry will be an infinitesimal point transformation 
if (and only if) $\tau$, $\xi$, $\psi$ have no dependence on $v$ and $\theta$,
or equivalently, $P^r$ and $P^\theta$ satisfy 
\begin{equation}\label{polar-X-point-cond}
P^r{}_{vv}=0, 
\quad
P^\theta{}_{\omega\omega}=0, 
\quad
P^r{}_\omega=0,
\quad
P^\theta{}_v=0 .
\end{equation}

To connect the invariance condition \eqref{polar-varX} to first integrals, 
consider the time derivative of a general first integral 
$I(t,r,\theta,v,\omega)$ of the central force \eom/ \eqref{polar-eom}. 
The chain rule combined with the determining equation \eqref{polar-I} 
directly yields the identity 
\begin{equation}\label{polar-I-id}
\frac{dI}{dt} = 
-\frac{\delta\L}{\delta r} I_v -\frac{\delta\L}{\delta \theta} r^{-2}I_\omega . 
\end{equation}
Note the variational derivatives will vanish precisely for 
solutions of the \eom/. 
The pair of coefficients of these derivatives are 
called the {\em multiplier} of the first integral, 
which we will denote 
\begin{equation}\label{polar-Q}
Q^r=-I_v,
\quad
Q^\theta = -r^{-2}I_\omega . 
\end{equation}
From determining equation \eqref{polar-I}, 
it is simple to see that there are no first integrals depending only on 
the variables $t$, $r$, $\theta$. 
Consequently, 
any two first integrals that differ by at most a constant 
will have the same multiplier, as given by equation \eqref{polar-Q}.  
Conversely, any multiplier determines a first integral 
to within an additive constant through inverting equation \eqref{polar-Q}
by a line integral, 
\begin{equation}\label{polar-Q-I}
I = \int_C \left( -Q^r dv -r^2 Q^\theta d\omega \right) 
\end{equation}
where $C$ denotes any curve in the coordinate space $(v,\omega)$,
starting at an arbitrary point and ending at a general point $(v,\omega)$. 

A comparison of identities \eqref{polar-I-id} and \eqref{polar-L-id}
combined with the invariance condition \eqref{polar-varX} 
now leads to the following general form of Noether's theorem.

\begin{prop}\label{polar-noether}
For the polar \eom/ \eqref{polar-eom}, 
variational symmetries in characteristic form 
$\hat\X=P^r\partial_r+P^\theta\partial_\theta$ 
and multipliers $(Q^r,Q^\theta)$ for first integrals 
have a one-to-one correspondence given by 
\begin{equation}\label{polar-P-Q}
P^r=Q^r, 
\quad
P^\theta=Q^\theta
\end{equation}
and 
\begin{equation}
R =I +\const
\end{equation}
In particular, through the relations \eqref{polar-Q} and \eqref{polar-P-Q}, 
variational point symmetries correspond to first integrals that are at most
quadratic in $v$ and $\omega$. 
\end{prop}

From this result, 
it is straightforward to obtain the symmetries that correspond to 
the first integrals \eqref{L}, \eqref{E}, \eqref{Theta}, \eqref{T} 
admitted by the polar \eom/. 

The first integrals \eqref{L} for angular momentum and \eqref{E} for energy 
have the respective multipliers 
\begin{equation}
(Q^r,Q^\theta)_{(L)}=(0,-1),
\quad
(Q^r,Q^\theta)_{(E)}=(-v,-\omega)
\end{equation}
which yield the infinitesimal symmetries (in characteristic form)
\begin{equation}\label{polar-L-E-Xhat}
\hat\X_{(L)} = -\partial_\theta,
\quad
\hat\X_{(E)} = -v\partial_r -\omega\partial_\theta . 
\end{equation}
From the relation \eqref{polar-Xhat},
these symmetries clearly correspond to infinitesimal point transformations 
\begin{equation}\label{polar-L-E-symm}
\X_{(L)}= -\partial_\theta,
\quad
\X_{(E)}= \partial_t . 
\end{equation}
Compared with the point transformations found earlier 
in \thmref{polar-pointsymm} for a general central force,
we see that $-\X_{(L)}=\X_1$ is the generator of 
a group of rotations \eqref{polar-rot}
and $\X_{(E)}=\X_2$ is the generator of a group of time-translations \eqref{polar-trans}. 

The multiplier of the first integral \eqref{Theta} 
for the angular quantity $\Theta$ is given by 
\begin{equation}
(Q^r,Q^\theta)_{(\Theta)} = (-v\Theta_{E},-\Theta_{L} -\omega\Theta_{E}) . 
\end{equation}
This yields the infinitesimal symmetry (in characteristic form)
\begin{equation}\label{polar-Theta-Xhat}
\hat\X_{(\Theta)} = -v\Theta_{E}\partial_r -(\Theta_{L} +\omega\Theta_{E})\partial_\theta . 
\end{equation}
Since $\Theta$ depends on $L$ and $E$ 
which themselves depend on the variables $v$ and $\omega$, 
the symmetry $\hat\X_{(\Theta)}$ is not an infinitesimal point transformation
(namely, its components do not satisfy condition \eqref{polar-X-point-cond}). 
However, 
this symmetry can be converted into a simpler form 
by putting $\tau= \Theta_{E}$ in the relations \eqref{polar-Xhat} and \eqref{polar-X}, 
giving the equivalent dynamical symmetry 
\begin{equation}\label{polar-Theta-symm}
\X_{(\Theta)} = \Theta_{E}\partial_t -\Theta_{L}\partial_\theta . 
\end{equation}

The first integral \eqref{T} for the temporal quantity $T$ 
has the analogous multiplier 
\begin{equation}
(Q^r,Q^\theta)_{(T)} = (-vT_{E},-T_{L} -\omega T_{E}) . 
\end{equation}
This yields the infinitesimal symmetry (in characteristic form)
\begin{equation}\label{polar-T-Xhat}
\hat\X_{(T)} = -vT_{E}\partial_r - (T_{L} +\omega T_{E})\partial_\theta
\end{equation}
which, similarly to $\hat\X_{(\Theta)}$, is not an infinitesimal point transformation
(since its components do not satisfy condition \eqref{polar-X-point-cond}). 
A simpler equivalent symmetry is obtained by putting $\tau= T_{E}$ 
in the relations \eqref{polar-Xhat} and \eqref{polar-X}, 
yielding the dynamical symmetry
\begin{equation}\label{polar-T-symm}
\X_{(T)} = T_{E}\partial_t -T_{L}\partial_\theta . 
\end{equation}

\begin{prop}
Through Noether's theorem for the polar \eom/ \eqref{polar-eom}, 
the first integrals for angular momentum \eqref{L} and energy \eqref{E}
correspond to the infinitesimal point symmetries \eqref{polar-L-E-symm}, 
while the first integrals given by the angular quantity \eqref{Theta} and the temporal quantity \eqref{T}
correspond respectively to the infinitesimal dynamical symmetries \eqref{polar-Theta-symm} and \eqref{polar-T-symm}.
\end{prop}

The infinitesimal dynamical symmetries given by $\X_{(\Theta)}$ and $\X_{(T)}$
each generate a group of transformations 
acting on the dynamical variables $(r(t),\theta(t),v(t),\omega(t))$
for any solution $(r(t),\theta(t))$ of the polar \eom/ \eqref{polar-eom}. 
We will see later that there is a simple way to derive the transformations
by first looking at how $\X_{(\Theta)}$ and $\X_{(T)}$ 
act on the first integrals $L$ and $E$. 
This will also lead to a simple way to find 
the structure of the four-dimensional group of transformations 
generated by all of the infinitesimal symmetries 
$\X_{(L)}$, $\X_{(E)}$, $\X_{(\Theta)}$,  and $\X_{(T)}$. 

\subsection{Transformation of first integrals under dynamical symmetries}

We now work out how the infinitesimal symmetries 
$\X_{(L)}$, $\X_{(E)}$, $\X_{(\Theta)}$, and $\X_{(T)}$
act on all of the first integrals $L$, $E$, $\Theta$, and $T$. 
Since every first integral is a function of 
the variables $t,r,\theta,v,\omega$,
the symmetries need to be prolonged to the coordinate space 
$(t,r,\theta,v,\omega)$. 
This is most easily carried out through the prolongation
relations \eqref{polar-prXrel} and \eqref{polar-prXhat} 
which use the equivalent symmetries 
$\hat\X_{(L)}$, $\hat\X_{(E)}$, $\hat\X_{(\Theta)}$, and $\hat\X_{(T)}$. 
For a general dynamical symmetry given by a generator \eqref{polar-X}, 
its action on a first integral $I(t,r,\theta,v,\omega)$ is thus given by 
\begin{equation}\label{prXhat-I}
\pr\X(I)\big|_\soln 
=\Big(\pr\hat\X(I) + \frac{dI}{dt}\Big)\big|_\soln
=\pr\hat\X|_\soln(I)
\end{equation}
which involves only the prolongation of the characteristic generator \eqref{polar-Xhat} evaluated on solutions of the polar \eom/ \eqref{polar-eom}.

In the case of the infinitesimal point symmetries $\X_{(L)}$ and $\X_{(E)}$,
the prolongation of their characteristic generators \eqref{polar-L-E-Xhat}
evaluated on the \eom/ \eqref{polar-eom} is straightforwardly given by 
\begin{equation}\label{polar-L-E-prXhat}
\pr\hat\X_{(L)}\big|_\soln 
= -\partial_\theta, 
\quad
\pr\hat\X_{(E)}\big|_\soln 
= -v\partial_r -\omega\partial_\theta -(\omega^2 r + F(r))\partial_v +(2 \omega v/r)\partial_\omega . 
\end{equation}
A direct calculation (using Maple) then yields
\begin{equation}
\pr\hat\X_{(L)}\big|_\soln (L) = \pr\hat\X_{(E)}\big|_\soln (L) = 0,
\quad
\pr\hat\X_{(L)}\big|_\soln (E) = \pr\hat\X_{(E)}\big|_\soln (E) =0 . 
\end{equation}
This result means that $L$ and $E$ can be treated as constants 
under the action of the symmetries $\hat\X_{(L)}$ and $\hat\X_{(E)}$. 
Hence the action of these symmetries on $\Theta$ and $T$ becomes 
simple to calculate, yielding the result  
\begin{equation}
\pr\hat\X_{(L)}\big|_\soln (T) = \pr\hat\X_{(E)}\big|_\soln (\Theta) = 0,
\quad
\pr\hat\X_{(L)}\big|_\soln (\Theta) = - \pr\hat\X_{(E)}\big|_\soln (T) = -1 . 
\end{equation}

The prolongation of 
the characteristic generators \eqref{polar-Theta-Xhat} and \eqref{polar-T-Xhat}
of the infinitesimal dynamical symmetries $\X_{(\Theta)}$ and $\X_{(T)}$ 
are more involved to calculate:
\begin{gather}
\begin{aligned}
\pr\hat\X_{(\Theta)}\big|_\soln = & 
\hat\X_{(\Theta)} 
+ \left(v\frac{d\Theta_{E}}{dt}\big|_\soln +(\omega^2 r + F(r))\Theta_{E}\right)\partial_v 
\\&\qquad
+ \left(\frac{d\Theta_{L}}{dt}\big|_\soln +\omega\frac{d\Theta_{E}}{dt}\big|_\soln-(2 \omega v/r)\Theta_{E}\right)\partial_\omega , 
\end{aligned}
\label{polar-Theta-prXhat}
\\
\begin{aligned}
\pr\hat\X_{(T)}\big|_\soln = & 
\hat\X_{(T)} 
+ \left(v\frac{dT_{E}}{dt}\big|_\soln +(\omega^2 r + F(r))T_{E}\right)\partial_v 
\\&\qquad
+ \left(\frac{dT_{L}}{dt}\big|_\soln +\omega\frac{dT_{E}}{dt}\big|_\soln -(2 \omega v/r)T_{E}\right)\partial_\omega . 
\end{aligned}
\label{polar-T-prXhat}
\end{gather}
One simplification is that, since $L$ and $E$ are constants of motion, 
the time-derivative terms can be calculated just by 
using the chain rule expression \eqref{polar-Dt}. 
This leads to the simple result (using Maple)
\begin{equation}
\pr\hat\X_{(\Theta)}\big|_\soln (E) = \pr\hat\X_{(T)}\big|_\soln (L) = 0,
\quad
\pr\hat\X_{(\Theta)}\big|_\soln (L) = -\pr\hat\X_{(T)}\big|_\soln (E) = 1 . 
\end{equation}
From these expressions, 
it is straightforward to calculate 
the action of the symmetries $\hat\X_{(\Theta)}$ and $\hat\X_{(T)}$ 
on $\Theta$ and $T$. 
The final result (using Maple) is again simple, 
\begin{equation}
\pr\hat\X_{(\Theta)}\big|_\soln (\Theta) = \pr\hat\X_{(T)}\big|_\soln (\Theta) = 0,
\quad
\pr\hat\X_{(\Theta)}\big|_\soln (T) = - \pr\hat\X_{(T)}\big|_\soln (T) = 1 . 
\end{equation}

By applying equation \eqref{prXhat-I} to the previous expressions, 
we now have the following useful result. 

\begin{thm}\label{polar-allI-transform}
For the first integrals $L$, $E$, $\Theta$, and $T$ 
of the polar \eom/ \eqref{polar-eom}, 
the action of the infinitesimal point symmetries $\X_{(L)}$, $\X_{(E)}$ 
consists of
\begin{gather}
\pr\X_{(L)}\big|_\soln (L) = \pr\X_{(L)}\big|_\soln (E) = \pr\X_{(L)}\big|_\soln (T) 
=0, 
\quad
\pr\X_{(L)}\big|_\soln (\Theta) = -1 , 
\\
\pr\X_{(E)}\big|_\soln (L) = \pr\X_{(E)}\big|_\soln (E) = \pr\X_{(E)}\big|_\soln (\Theta) = 0,
\quad
\pr\X_{(E)}\big|_\soln (T) =1 , 
\end{gather}
and the action of the infinitesimal dynamical symmetries $\X_{(\Theta)}$ and $\X_{(T)}$
consists of 
\begin{gather}
\pr\X_{(\Theta)}\big|_\soln (E) = \pr\X_{(\Theta)}\big|_\soln (\Theta) = \pr\X_{(\Theta)}\big|_\soln (T) = 0,
\quad
\pr\X_{(\Theta)}\big|_\soln (L) =1 , 
\\
\pr\X_{(T)}\big|_\soln (L) = \pr\X_{(T)}\big|_\soln (\Theta) = \pr\X_{(T)}\big|_\soln (T) = 0,
\quad
\pr\X_{(T)}\big|_\soln (E) = -1 . 
\end{gather}
Hence the four first integrals are canonical coordinates for these four infinitesimal symmetries. 
\end{thm}

We will next use this result to give a simple derivation of 
the group of transformations generated by each of 
the infinitesimal dynamical symmetries $\X_{(\Theta)}$ and $\X_{(T)}$. 

The components of these two symmetries are functions only of $r$, $L$ and $E$,
where $L$ and $E$ are the first integrals \eqref{L} and \eqref{E}.
Hence, the group of transformations can be found by integrating 
a system of equations involving only the basic variables $t$, $r$, $\theta$,
along with $L$ and $E$ regarded as auxiliary variables. 
From the expression \eqref{polar-Theta-symm} for $\X_{(\Theta)}$, 
this system is given by 
\begin{gather}
\parder{t(\epsilon)}{\epsilon} 
= \Theta_{E}(r(\epsilon),L(\epsilon),E(\epsilon)),
\quad
\parder{\theta(\epsilon)}{\epsilon} 
= \Theta_{L}(r(\epsilon),L(\epsilon),E(\epsilon))
\label{polar-Theta-groupeqn}\\
\parder{r(\epsilon)}{\epsilon} = 0,
\quad
\parder{L(\epsilon)}{\epsilon} = 1,
\quad
\parder{E(\epsilon)}{\epsilon} = 0
\label{polar-Theta-inv-eqn}
\end{gather}
where $\epsilon$ is the group parameter. 
Integration of equation \eqref{polar-Theta-inv-eqn} easily gives 
\begin{equation}\label{polar-Theta-auxgroup}
r(\epsilon) = r,
\quad
L(\epsilon) = L+\epsilon,
\quad
E(\epsilon) = E . 
\end{equation}
Equation \eqref{polar-Theta-groupeqn} then becomes
\begin{gather}
\parder{\theta(\epsilon)}{\epsilon} 
= \Theta_{L}(r,L+\epsilon,E)
= -\parder{\Theta_{0}(r,L+\epsilon,E)}{\epsilon} , 
\\
\begin{aligned}
\parder{t(\epsilon)}{\epsilon} 
= \Theta_{E}(r,L+\epsilon,E)
& = -\frac{(L+\epsilon)\Theta_{L}(r,L+\epsilon,E)}{2(E-U(r))} 
\\
&= -\parder{\Theta_{0}(r,L+\epsilon,E)}{\epsilon} \frac{L+\epsilon}{2(E-U(r))} ,
\end{aligned}
\end{gather}
which can be straightforwardly integrated. 
This yields
\begin{equation}\label{polar-Theta-group}
\begin{aligned}
& \theta(\epsilon) 
=\theta + \Theta_{0}(r,L,E) - \Theta_{0}(r,L+\epsilon,E)
= \Theta - \Theta_{0}(r,L+\epsilon,E)
\\
& t(\epsilon)
= t +\frac{L( \Theta_{0}(r,L,E) - \Theta_{0}(r,L+\epsilon,E) )}{2(E-U(r))}
\end{aligned}
\end{equation}
which is the transformation group generated by $\X_{(\Theta)}$. 
Similar steps applied to the expression \eqref{polar-T-symm} for $\X_{(T)}$
leads to the transformation group
\begin{equation}\label{polar-T-group}
\begin{aligned}
& t(\epsilon) 
=t + T_{0}(r,L,E) - T_{0}(r,L,E-\epsilon)
= T - T_{0}(r,L,E-\epsilon)
\\
& \theta(\epsilon)
=\theta +\frac{r^2( T_{0}(r,L,E) - T_{0}(r,L,E-\epsilon) )}{L}
\end{aligned}
\end{equation}
with 
\begin{equation}\label{polar-T-auxgroup}
r(\epsilon) = r,
\quad
L(\epsilon) = L
\quad
E(\epsilon) = E-\epsilon . 
\end{equation}
Recall that the respective transformation groups 
generated by $\X_{(L)}$ and $\X_{(E)}$ are simply 
\begin{gather}
\theta(\epsilon) =\theta -\epsilon 
\label{polar-L-group}\\
t(\epsilon) =t + \epsilon
\label{polar-E-group}
\end{gather}
with 
\begin{equation}\label{polar-L-E-auxgroup}
r(\epsilon) = r,
\quad
L(\epsilon) = L
\quad
E(\epsilon) = E . 
\end{equation}

Finally, we can use the same approach to obtain the Lie algebra generated by 
the infinitesimal dynamical symmetries $\X_{(\Theta)}$ and $\X_{(T)}$,
together with the infinitesimal point symmetries $\X_{(L)}$ and $\X_{(E)}$. 
Through \thmref{polar-allI-transform},
we extend these symmetries to act on $L$ and $E$ as auxiliary variables:
\begin{gather}
\X_{(L)}^\ext= -\partial_\theta, 
\quad
\X_{(E)}^\ext= \partial_t, 
\label{X(L)X(E)}\\
\X_{(\Theta)}^\ext = \Theta_{E}\partial_t -\Theta_{L}\partial_\theta + \partial_L,
\quad
\X_{(T)}^\ext = T_{E}\partial_t -T_{L}\partial_\theta - \partial_E . 
\label{X(Theta)X(T)}
\end{gather}
This extension can be viewed as a prolongation that is restricted to the variables $(t,r,\theta,L,E)$ appearing in the first integrals. 

The extended generators \eqref{X(L)X(E)}--\eqref{X(Theta)X(T)} have the form of infinitesimal point transformations on the space $(t,r,\theta,L,E)$. 
A straightforward calculation (using Maple) 
shows that the commutators of these generators vanish. 
Thus, we have the following new result. 

\begin{thm}\label{polar-symmgroup}
For the polar \eom/ \eqref{polar-eom}, 
the the infinitesimal dynamical symmetries $\X_{(\Theta)}$ and $\X_{(T)}$, 
and infinitesimal point symmetries $\X_{(L)}$ and $\X_{(E)}$,
together generate a four-dimensional abelian group of transformations 
\eqref{polar-Theta-group}, \eqref{polar-T-group}, 
\eqref{polar-L-group}, \eqref{polar-E-group}. 
When these transformations are extended in the form 
\eqref{polar-Theta-auxgroup}, \eqref{polar-T-auxgroup}, 
and \eqref{polar-L-E-auxgroup},
then the extended group acts as commuting point transformations on 
$(t,r,\theta,L,E)$ under which $r$ is invariant. 
\end{thm}

\subsection{Method of extended point symmetries for finding first integrals and hidden dynamical symmetries}

The results in \thmref{polar-symmgroup} and \thmref{polar-allI-transform}
have the important consequence that we can now formulate 
an algorithmic symmetry method to obtain all of the first integrals 
along with their underlying hidden symmetry group 
by using only point symmetries combined with Noether's theorem. 
The method has four steps. 

First, 
we find the variational point symmetries of the polar \eom/ \eqref{polar-eom},
and through Noether's theorem, 
we obtain the corresponding first integrals.
This will give $\X=\partial_\theta$ and $\X=\partial_t$,
yielding $L$ and $E$. 
For later use, we also identify the joint invariant of these point symmetries 
in the space $(t,r,\theta)$,
which is $r$. 

Second, 
we re-write the polar \eom/ in the form of an equivalent first-order system 
using only the basic dynamical variables $r$, $\theta$, 
plus the two first integrals, $L$, $E$ 
(after elimination of $v$ and $\omega$ in terms of $L$ and $E$). 
This system consists of the dynamical equations 
\begin{equation}\label{1storderdynsystem}
\frac{dL}{dt}= 0,
\quad
\frac{dE}{dt}= 0,
\quad
\frac{d\theta}{dt}= r^{-2}L, 
\quad
\frac{dr}{dt}= \pm\sqrt{2E-r^{-2}L^2-2U(r)} . 
\end{equation} 

Third, we find point symmetries of this first-order dynamical system 
under the restrictions that they commute with the variational point symmetries of the original \eom/ and that they preserve the joint invariant of those symmetries. 
The form of the symmetry generators being sought is thus 
\begin{equation}\label{specialdynsymm}
\X=\eta^t(r,L,E)\partial_t + \eta^\theta(r,L,E)\partial_\theta + \eta^L(r,L,E)\partial_L + \eta^E(r,L,E)\partial_E . 
\end{equation} 
This will give the four symmetries \eqref{X(L)X(E)}, \eqref{X(Theta)X(T)}, 
up to expressions which are arbitrary functions of $L$ and $E$. 

Last, we find the canonical coordinates of these point symmetries 
in the space $(t,r,\theta,L,E)$. 
This will reproduce the first integrals $L$ and $E$ as well as yield the two additional first integrals $\Theta$ and $T$. 

It is important to emphasize that the determining equations for point symmetries
with the special form \eqref{specialdynsymm} will be a linear system of four equations in four unknowns,
whose solution will not involve the first integrals $\Theta$ and $T$. 
In particular, the system simply consists of 
\begin{gather}
\partial_r\eta^L=0,
\quad
\partial_r\eta^E=0,
\\
\partial_r\eta^t= \frac{r^{-2}L\eta^L-\eta^E}{(2(E-U(r))-r^{-2}L^2)^{3/2}},
\quad
\partial_r\eta^\theta= \frac{r^{-2}(2(E-U(r))\eta^E-L\eta^L)}{(2(E-U(r))-r^{-2}L^2)^{3/2}} . 
\end{gather}
This is in contrast to the case of general point symmetries,
where the determining equations will be an underdetermined system 
whose solution will necessarily contain expressions which are arbitrary functions of all of the first integrals.

\section{First integrals in $n$ dimensions}
\label{ndim}

We return to the central force \eom/ in the $n$-dimensional form \eqref{ndim-sys} given in terms of the position vector $\rvec$ and velocity vector $\vvec$
in $\Rnum^n$. 
Hereafter, we will put $m=1$ without loss of generality
(via rescaling the physical units of the dynamical variables). 

We recall that all solutions $(\rvec(t),\vvec(t))$ of the \eom/ lie in a time-independent plane spanned by these two vectors. 
Let $\rhatvec$ be the unit vector along $\rvec$ 
and $\thhatvec$ be the unit vector orthogonal to $\rvec$ in the plane of motion,
such that $\{\rhatvec,\thhatvec\}$ is a right-handed orthonormal basis in this plane $\Rnum^2\subset \Rnum^n$. 
Then the vectors $\rvec$ and $\vvec$ are related to the polar variables $(r,\theta,v,\omega)$ by 
\begin{equation}\label{ndim-rvec-vvec}
\rvec = r\rhatvec,
\quad
\vvec = v\rhatvec + r\omega\thhatvec
\end{equation}
and 
\begin{equation}\label{e1e2}
\hat e_1 = \cos(\theta)\rhatvec -\sin(\theta)\thhatvec, 
\quad
\hat e_2 = \sin(\theta)\rhatvec +\cos(\theta)\thhatvec
\end{equation}
where $\{\hat e_1,\hat e_2\}$ is a fixed (time-independent) orthonormal basis 
aligned with the respective directions given by $\theta=0$ and $\theta=\pi/2$ 
in the plane of motion. 
Thus we have
\begin{equation}\label{rv-rels}
r=|\rvec|=\rhatvec\cdot\rvec,
\quad
v=\rhatvec\cdot\vvec,
\quad
\omega = r^{-1}\thhatvec\cdot\vvec
\end{equation}
and 
\begin{equation}
\cos(\theta) =\rhatvec\cdot\hat e_1 = \thhatvec\cdot\hat e_2,
\quad
\sin(\theta) =\rhatvec\cdot\hat e_2 = -\thhatvec\cdot\hat e_1 . 
\end{equation}

From the \eom/ \eqref{ndim-sys} for $(\rvec,\vvec)$, 
we easily derive 
\begin{equation}
\frac{d\rhatvec}{dt} = \omega \thhatvec,
\quad
\frac{d\thhatvec}{dt} = -\omega \rhatvec
\end{equation}
which implies that the antisymmetric product of $\rhatvec$ and $\thhatvec$ is time-independent 
\begin{equation}
\frac{d}{dt}(\rhatvec\wedge\thhatvec)=0 . 
\end{equation}
Thus
\begin{equation}
\Lhattens = \rhatvec\wedge\thhatvec 
\end{equation}
is an antisymmetric tensor (bi-vector) in $\Rnum^n$, 
satisfying 
\begin{equation}
\frac{d\Lhattens}{dt} =0, 
\quad
\Lhattens\wedge\Lhattens=0,
\quad
\Lhattens = \hat e_1\wedge \hat e_2
\end{equation}
and
\begin{equation}\label{Lhattens-rvec-thvec}
\rhatvec\cdot\Lhattens = \thhatvec,
\quad
\thhatvec\cdot\Lhattens = -\rhatvec,
\quad
\Lhattens\cdot\Lhattens = 2 . 
\end{equation}
There is a one-to-one correspondence between this tensor and the plane of motion. 
It follows that $\Lhattens$ is a first integral of the $n$-dimensional \eom/
and determines the orientation of the plane of motion. 
Note this first integral does not appear among the polar first integrals 
since it has no dynamical content for the motion within that plane.

Using these preliminaries, we now will express the polar first integrals \eqref{L}, \eqref{E}, \eqref{Theta}, \eqref{T} in a geometric $n$-dimensional form. 

We start with the energy \eqref{E}. 
We re-write this first integral by noting 
\begin{equation}\label{ndim-vsq}
|\vvec|^2 = v^2 + L^2/r^2
\end{equation}
so thus 
\begin{equation}\label{ndim-E}
E= \tfrac{1}{2}|\vvec|^2 + U(|\rvec|)
\end{equation}
is directly in an $n$-dimensional form. 

Next we consider the angular momentum \eqref{L}. 
This first integral can be written as 
\begin{equation}\label{ndim-L}
L= |\rvec| \thhatvec\cdot\vvec
\end{equation}
by using the vector $\thhatvec$. 
A more physically and geometrically natural formulation is given by 
combining $L$ and $\Lhattens$ into the first integral 
\begin{equation}\label{ndim-Lhattens}
\Ltens = L\Lhattens = L\rhatvec\wedge\thhatvec . 
\end{equation}
From relation \eqref{ndim-rvec-vvec}, we can express
\begin{equation}\label{ndim-thvec}
\thhatvec= L^{-1}( |\rvec|\vvec - (\vvec\cdot\rvec)\rhatvec )
\end{equation}
in terms of $\rvec$ and $\vvec$, 
yielding 
\begin{equation}\label{ndim-Ltens}
\Ltens = \rvec\wedge\vvec . 
\end{equation}
This first integral is an antisymmetric tensor (bi-vector) in $\Rnum^n$ having the properties
\begin{equation}
\frac{d\Ltens}{dt} =0, 
\quad
|\Ltens|^2 = 2|L|^2,
\quad
\Ltens\wedge\Ltens=0 . 
\end{equation}

We next turn to the angular quantity \eqref{Theta}, 
which can be expressed as 
\begin{equation}\label{theta+Theta0}
\Theta = \theta + \Phi
\end{equation}
in terms of the integral expression 
\begin{equation}\label{Theta0}
\Phi = - L\int^{|\rvec|}_{r_0} \frac{\sgn(v)}{\sqrt{ 2(E+U(r_\eq)-U(r))r^4 -L^2 r^2}}\,dr .
\end{equation}
Since $\Theta$ is an angle in the plane of motion in $\Rnum^n$, 
it can be geometrically formulated as a unit vector 
\begin{equation}\label{ndim-unitvecTheta}
\Thvec = \cos(\Theta)\hat e_1 + \sin(\Theta)\hat e_2 . 
\end{equation}
This vector is a first integral and has the properties
\begin{equation}
\frac{d\Thvec}{dt}=0,
\quad
|\Thvec|=1,
\quad
\Thvec\wedge\Ltens =0 . 
\end{equation}
We can write it in terms of the unit vectors $\rhatvec$ and $\thhatvec$, 
or equivalently the position and velocity vectors $\rvec$ and $\vvec$, 
by substituting the expressions \eqref{theta+Theta0} and \eqref{e1e2}, 
followed by using the relations \eqref{rv-rels} and \eqref{ndim-thvec}, 
which gives
\begin{equation}\label{ndim-Theta}
\Thvec = \cos(\Phi)\rhatvec + \sin(\Phi)\thhatvec
= \frac{1}{|\rvec|}\Big(\cos(\Phi) -\frac{\rvec\cdot\vvec}{L}\sin(\Phi)\Big)\rvec + \frac{|\rvec|}{L}\sin(\Phi)\vvec
\end{equation}
where the integral expression $\Phi$ involves $L$, $E$, $|\rvec|$, 
and also $\sgn(v)$. 
Note the dependence on $\sgn(v)$ drops out of $\cos(\Phi)=\cos|\Phi|$
(since it is an even function). 
It is useful to re-write $-\sin(\Phi)=\sgn(v)\sgn(L)\sin|\Phi|$ 
so that $\sgn(v)$ appears as an overall factor. 
After re-writing these terms in the expression $\cos(\Phi)\rhatvec + \sin(\Phi)\thhatvec$, 
and using the relation \eqref{Lhattens-rvec-thvec}, 
we obtain 
\begin{equation}\label{ndim-Thetavec}
\Thvec = |\rvec|^{-1}\left( \cos(\phi_0) \rvec -\sqrt{2}\sgn(\vvec\cdot\rvec)|\Ltens|^{-1}\sin(\phi_0) \rvec\cdot\Ltens \right)
\end{equation}
which has a geometrically simple form,
where
\begin{equation}\label{phi}
\phi_0=|\Phi| 
= |\Ltens|\int^{|\rvec|}_{r_0} \sqrt{ 4(E+U(r_\eq)-U(r))r^4 -|\Ltens|^2 r^2}^{\,-1}\,dr
\end{equation}
is a time-dependent angle. 
Here $r_\eq$ is an equilibrium point \eqref{equilpoint} of the central force potential $U(|\rvec|)$, 
and $r_0$ is a turning point \eqref{turningpoint} or an inertial point \eqref{inertialpoint} of the effective potential 
\begin{equation}\label{ndim-Ueff}
U_\eff(|\rvec|) = |\Ltens|^2/(2|\rvec|)^2 + U(|\rvec|) -U(r_\eq) . 
\end{equation}

We remark that another vector first integral can be formed from $\Theta$ 
by rotating $\Thvec$ by the angle $\pi/2$ in the plane of motion. 
This yields 
\begin{equation}\label{ndim-unitvecThetaperp}
\Thvec^\perp = -\sin(\Theta)\hat e_1 + \cos(\Theta)\hat e_2
= -\sin(\Phi)\rhatvec+ \cos(\Phi)\thhatvec 
\end{equation}
which we can write equivalently in the geometrical form 
\begin{equation}\label{ndim-Thetavecperp}
\Thvec^\perp = \Thvec\cdot\Lhattens 
= (|\rvec|L)^{-1}\left( \cos(\phi_0) \rvec\cdot\Ltens +\tfrac{1}{\sqrt{2}}\sgn(\vvec\cdot\rvec)|\Ltens|\sin(\phi_0) \rvec \right) . 
\end{equation}
It has the properties 
\begin{equation}
\frac{d\Thvec^\perp}{dt}=0,
\quad
|\Thvec^\perp|=1,
\quad
\Thvec^\perp\wedge\Ltens =0,
\quad
\Thvec^\perp\cdot\Thvec =0 . 
\end{equation}

Finally, we consider the temporal quantity \eqref{T}. 
This first integral has the $n$-dimensional form
\begin{equation}\label{ndim-T}
T = t- \sqrt{2}\sgn(\vvec\cdot\rvec)\tau_0
\end{equation}
with 
\begin{equation}\label{tau}
\tau_0 = \int^{|\rvec|}_{r_0} \sqrt{ 4(E+U(r_\eq)-U(r)) -|\Ltens|^2 r^{-2}}^{\,-1}\,dr 
\end{equation}
which we obtain through the relations \eqref{rv-rels}. 
Note $\tau_0$ is a time-dependent geometrical expression,
where $r_\eq$ and $r_0$ are the same radial points used in the first integral $\Thvec$. 

In summary, 
the four polar first integrals \eqref{L}, \eqref{E}, \eqref{Theta}, \eqref{T} 
give rise to corresponding $n$-dimensional first integrals 
\eqref{ndim-Ltens}, \eqref{ndim-E}, \eqref{ndim-Thetavec}, \eqref{ndim-T} 
which each have a geometrical form given in terms of the position and velocity vectors $\rvec$ and $\vvec$. 
Note that the energy \eqref{E} and the temporal quantity \eqref{T} 
are scalar first integrals, 
whereas the angular momentum \eqref{ndim-Ltens} is a tensor first integral
and the angular quantity \eqref{ndim-Thetavec} is a vector first integral. 
We will show later that this vector first integral can be used to define 
a general \LRL/ vector. 

The total number of time-independent quantities defined by all of these $n$-dimensional first integrals is $n(n-1)/2$ from the components of $\Ltens$, 
and $n$ from the components of $\Thvec$
(as defined with respect to any fixed orthonormal basis of $\Rnum^n$), 
plus $2$ given by $E$ and $T$. 
However, not all of these quantities are independent, 
due to the algebraic properties 
$\Ltens\wedge\Ltens=0$, $\Thvec\wedge\Ltens =0$, $|\Thvec|=1$. 
In particular, 
$\Ltens$ has only $2n-3$ independent components, 
corresponding to the orientation of the $2$-dimensional plane of motion in $\Rnum^n$ 
(which accounts for $2n-4$ components)
and the magnitude of the angular momentum in this plane
(which accounts for a single component);
$\Thvec$ has $n-1$ independent components, 
corresponding to a $2$-dimensional cone in $\Rnum^n$
(which accounts for $n-2$ components)
and an angle of a unit vector in this cone
(which accounts for a single component),
where the cone is tangent to the plane of motion. 
Consequently, 
together $\Ltens$ and $\Thvec$ determine a total of $2n-2$ independent quantities.
Since there are $2$ independent quantities given by $E$ and $T$, 
this yields $2n$ independent quantities altogether,  
which is precisely the number of functionally independent first integrals 
allowed for a second-order system with $n$ independent dynamical variables. 

Thus we have the following result. 

\begin{thm}\label{ndim-allI}
For the \eom/ \eqref{ndim-eom} of general central force dynamics
in $n>1$ dimensions:
\newline 
{\rm (1)}
$\Ltens$ and $E$ are well-defined first integrals for all solutions $\rvec(t)$. 
\newline 
{\rm (2)}
$\Thvec$ and $T$ are well-defined first integrals for all non-circular solutions $\rvec(t)$. 
\newline 
{\rm (3)}
Altogether, $\Ltens$, $E$, $\Thvec$ yield $2n-1$ functionally independent \coms/, while $T$ is a first integral which is not a \com/. 
\newline 
{\rm (4)}
Every first integral is a function of $\Ltens$, $E$, $\Thvec$, $T$,
and every \com/ is a function of $\Ltens$, $E$, $\Thvec$.  
\end{thm}

We emphasize that each of $\Ltens$, $E$, $\Thvec$, $T$ 
can be directly verified to obey the defining equation \eqref{ndim-1stintegr}
for first integrals of the central force \eom/ \eqref{ndim-eom}. 
We also emphasize that the integral expressions \eqref{phi} and \eqref{tau}
appearing respectively in $\Ltens$ and $T$
are well-defined because we have specified the endpoint value $r_0$ 
directly in terms of the effective potential \eqref{ndim-Ueff}. 
If the endpoint were omitted or allowed to be arbitrary, 
then the resulting first integrals would be defined only to within an arbitrary additive constant and would lose an important part of their physical meaning, which we will discuss next. 

\subsection{General \LRL/ vector}

The vector first integral $\Thvec$ has physical properties 
similar to the angular first integral $\Theta$ discussed in \secref{interpret}. 
These properties depend on the particular choice made for the radial endpoint $r_0$ in the integral expression \eqref{phi} appearing in $\Thvec$. 

There are two different possibilities for $r_0$, 
which are tied to the shape of trajectories in the plane of motion
for solutions $\rvec(t)$ of the central force \eom/ \eqref{ndim-eom}. 
Recall, for a given non-circular trajectory, 
a turning point is a radial value at which the radius $|\rvec|$ of the trajectory is a local extremum (\ie/ an apsis), 
and an inertial point is a radial value at which the radial speed $\rhatvec\cdot\vvec$ is a local extremum. 
For all central force potentials $U(|\rvec|)$ of physical interest, 
at least one turning point or one inertial point can be assumed to exist
on each non-circular trajectory. 

When a trajectory has no turning points, 
then we choose $r_0$ to be the radius of an inertial point on the trajectory in the plane of motion. 
In this case, $\Thvec$ will be the direction vector of the radial line that intersects the trajectory at this point. 
Similarly, the first integral $T$ will be the time $t$ at which the trajectory reaches this point. 

When a trajectory instead has at least one turning point, 
then we can choose $r_0$ to be this turning point on the trajectory in the plane of motion. 
Then $\Thvec$ will be the direction vector of the radial line that intersects the trajectory at this point,
and the first integral $T$ will be the time $t$ at which the trajectory reaches this point. 
If the trajectory possesses a single turning point, 
this point will be a periapsis point 
(\ie/ a local minimum of the radius $|\rvec|$),
and $\Thvec$ is then uniquely defined. 
In contrast, if the trajectory possesses multiple turning points,
then the trajectory will have apses that occur in pairs consisting of 
a periapsis point and an apoapsis point 
(\ie/ a local maximum of the radius $|\rvec|$). 
In this case, 
$\Thvec$ will be single-valued only on each part of the trajectory 
between pairs of successive periapsis points or successive apoapsis points.
This is caused by the sign factor $\sgn(\vvec\cdot\rvec)$ appearing in $\Thvec$,
which produces a jump in the value of $\Thvec$ when the trajectory passes through the next apsis point occurring after the apsis point that corresponds to the radius $r_0$. 
From \propref{apsisangle}, 
this jump is equal to the angular separation \eqref{precessangle} 
between each pair successive periapsis points or successive apoapsis points on the trajectory. 

If this angular separation \eqref{precessangle} is a rational multiple of $2\pi$, 
then the shape of the trajectory is closed. 
In this case the first integral $\Thvec$ will yield a finite number of distinct vectors in the plane of motion, 
corresponding to the finite number of periapsis points or apoapsis points.
When the angular separation is exactly $2\pi$, 
there will be exactly two apsis points 
and then the trajectory is not be precessing in the plane of motion. 
Otherwise, there will be more than two apsis points, 
and then the trajectory is precessing. 

In contrast, 
if the angular separation \eqref{precessangle} is an irrational multiple of $2\pi$, 
then the curve describing the shape of the trajectory is open. 
In this case the first integral $\Thvec$ will yield an infinite number of distinct vectors in the plane of motion,
corresponding to the infinite number of periapsis points or apoapsis points,
and the trajectory is then precessing in the plane of motion. 

In either case, 
the temporal first integral $T$ yields a periodic infinite sequence of values,
which are the times at which the trajectory passes through successive periapsis or apoapsis points. 

Based on these properties,
it is natural to view the vector $\Thvec$ as defining the directional part of a general \LRL/ vector for arbitrary central forces in $n>1$ dimensions. 
We first re-write the expression \eqref{ndim-Theta} for $\Thvec$ in the following way, 
which will make contact with the usual \LRL/ vector expression 
for an inverse-square central force \cite{GolPooSaf,Fra,Per}. 

From expression \eqref{phi}, we have the relation
$\partial_{|\rvec|} \cos(\phi_0) = -|L||\rvec|^{-1}|\vvec\cdot\rvec|^{-1}\sin(\phi_0)$
which yields
\begin{equation}\label{rel1}
\sgn(\vvec\cdot\rvec)\sin(\phi_0) = - \frac{|\rvec|\vvec\cdot\rvec}{|L|} \partial_{|\rvec|} \cos(\phi_0) . 
\end{equation}
By using expressions \eqref{ndim-rvec-vvec} and \eqref{ndim-Lhattens}, 
we also get the relation
\begin{equation}\label{rel2}
(\vvec\cdot\rvec)\rhatvec\cdot\Ltens = |\rvec|\vvec\cdot\Ltens  + L^2\rhatvec . 
\end{equation}
These relations \eqref{rel2} and \eqref{rel1} can be combined with expression \eqref{ndim-Thetavec}, 
yielding
\begin{equation}\label{ndim-Thetavec-alt}
\Thvec = \frac{\partial_{|\rvec|}(|\rvec|\cos(\phi_0))}{|\rvec|} \rvec + \frac{2|\rvec|^2\partial_{|\rvec|}(\cos(\phi_0))}{|\Ltens|^2} \vvec\cdot \Ltens . 
\end{equation}

Thus, we obtain the general \LRL/ vector 
\begin{equation}\label{ndim-LRLvec}
\LRLvec = A(E,L) \Big( \frac{\partial_{|\rvec|}(|\rvec|\cos(\phi_0))}{|\rvec|} \rvec + \frac{2|\rvec|^2\partial_{|\rvec|}(\cos(\phi_0))}{|\Ltens|^2} \vvec\cdot \Ltens \Big)
\end{equation}
where $A(E,L)>0$ is an arbitrary normalization factor,
and $\phi_0$ is the (time-dependent) angle expression \eqref{phi}. 
This vector satisfies 
\begin{equation}
\frac{d\LRLvec}{dt}=0,
\quad
\LRLvec\wedge\Ltens =0 . 
\end{equation}
From the preceding discussion, we have the following main result. 

\begin{thm}
For all non-circular solutions of the \eom/ \eqref{ndim-eom} of 
general central force dynamics in $n>1$ dimensions, 
the vector quantity \eqref{ndim-LRLvec} is a well-defined first integral $\LRLvec$ 
whose geometrical properties are determined by 
the radial value $r_0>0$ appearing in the angular quantity $\phi_0$. 
\newline
{\rm (1)}
 $\LRLvec$ lies along the direction of the radial vector $\rvec(T)$ 
in the plane of motion of the trajectory $\rvec(t)$,
where $T$ is the temporal first integral \eqref{ndim-T} 
using the radial value $r_0$. 
\newline
{\rm(2)}
For an unbounded trajectory $\rvec(t)$ with no radial turning points, 
$T$ yields the times $t$ when $|\rvec(t)|=r_0$ is the radius of the two inertial points on the trajectory 
(where the radial speed $\vvec(t)\cdot\rhatvec$ is a positive or negative extremum);
thereby $\LRLvec$ is double-valued. 
\newline
{\rm(3)}
For an unbounded trajectory $\rvec(t)$ with one radial turning point, 
$T$ yields the time $t$ when $|\rvec(t)|=r_0$ is the radius of the periapsis point on the trajectory 
(where the radius $|\rvec(t)|$ is a minimum);
thereby $\LRLvec$ is single-valued. 
\newline
{\rm (4)}
For a bounded trajectory $\rvec(t)$ with multiple radial turning points, 
$T$ yields the times $t$ when $|\rvec(t)|=r_0$ is the radius of either the periapsis point(s) or the apoapsis point(s) on the trajectory 
(where the radius $|\rvec(t)|$ is, respectively, either a minimum or a maximum);
therefore, 
$\LRLvec$ is single-valued if the angular separation between successive 
periapsis and apoapsis points is $\pi$, and otherwise $\LRLvec$ is multi-valued.
\end{thm}

An interesting variant of the general \LRL/ vector $\LRLvec$
can be obtained by choosing the radial value $r_0$ to be the radius of an inertial point for any non-circular trajectory. 
As we will see in the next section, 
this variant generalizes Hamilton's eccentricity vector \cite{Cor} 
which is known to be a first integral for inverse-square central forces.

\section{Examples of $n$-dimensional general \LRL/ vector}
\label{LRLexamples}

We will now examine the general \LRL/ vector \eqref{ndim-LRLvec} and its variant
for two important examples of central force dynamics in $n$ dimensions,
using the results from \secref{examples}. 

For comparison with previous results in the literature on central force dynamics in $n=3$ dimensions, 
it will be useful to note some 3-dimensional identities relating 
the antisymmetric product of vectors and the cross-product of vectors:
\begin{gather}
\vec C\cdot(\vec A\wedge\vec B) = (\vec C\cdot\vec A)\vec B - (\vec C\cdot\vec B)\vec A = (\vec A\times\vec B)\times\vec C
\\
(\vec A\wedge\vec B)\cdot(\vec C\wedge\vec D) = 2(\vec A\cdot\vec B)(\vec C\cdot\vec D)- 2(\vec A\cdot\vec D)(\vec B\cdot\vec C) = 2(\vec A\times\vec B)\cdot(\vec C\times\vec D)
\end{gather}

\subsection{Inverse-square force}
An inverse-square force \eqref{invsqF} has 
three different types of non-radial trajectories, 
depending on the first integrals \eqref{ndim-Ltens} and \eqref{ndim-E}
for angular momentum $\Ltens$ and energy $E$:
elliptic trajectories, with $0>E>E_\min$;
parabolic trajectories, with $E=0$;
and hyperbolic trajectories, with $E>0$.

Parabolic and hyperbolic trajectories respectively have 
a single turning point \eqref{kepler-E=0-tp} and \eqref{kepler-posE-tp},
which is the periapsis on the trajectory. 
Using the radius of the turning point for the radial value $r_0$ 
to evaluate the angular expression \eqref{Theta0}, we have 
\begin{equation}
\Phi= \arctan\left(\frac{L^2-kr}{Lvr}\right) -\sgn(vL) \frac{\pi}{2}
\end{equation}
where $L=\tfrac{1}{2}\Ltens\cdot\Lhattens$ is the angular momentum scalar 
and $v=\vvec\cdot\rhatvec$ is the radial speed. 
This yields 
\begin{equation}
\cos(\Phi) = \frac{L^2-kr}{r\sqrt{2EL^2+k^2}}=\cos(\phi_0),
\quad
\sin(\Phi) = \frac{-Lv}{\sqrt{2EL^2+k^2}}
\end{equation}
Hence the vector first integral \eqref{ndim-Thetavec-alt} is given by 
\begin{equation}\label{kelper-posE-E=0-Thvec}
\Thvec 
= \frac{1}{r^2\sqrt{2EL^2+k^2}}\left( (L^2-kr)\rvec - (\vvec\cdot\rvec)\rvec\cdot\Ltens \right)
= \frac{-1}{\sqrt{E|\Ltens|^2+k^2}}\left( (k/|\rvec|)\rvec +\vvec\cdot\Ltens \right)
\end{equation}
in terms of the energy $E$ and the angular momentum $\Ltens$. 
The expression \eqref{ndim-LRLvec} for the general \LRL/ vector thereby yields
\begin{equation}\label{kelper-posE-E=0-LRLvec}
\LRLvec_* = -\left( (k/|\rvec|)\rvec +\vvec\cdot\Ltens \right)
\end{equation}
where we have chosen the normalization factor to be 
$A(E,L)=\sqrt{E|\Ltens|^2+k^2}$. 
(Note the subscript $*$ indicates the use of a turning point for $r_0$
in defining the vector.)
This vector \eqref{kelper-posE-E=0-LRLvec} lies along 
the radial line connecting the origin to the periapsis point on the trajectory. 

Elliptic trajectories have two turning points \eqref{kepler-negE-tp},
which are the periapsis and apoapsis on the trajectory. 
Evaluating the angular expression \eqref{Theta0} 
by using the radius $r_{*\pm}$ of the turning points for the radial value $r_0$, 
we obtain 
\begin{equation}
\Phi_\pm= \arctan\left(\frac{L^2-kr}{Lvr}\right)\pm \sgn(vL)\frac{\pi}{2} . 
\end{equation}
This yields 
\begin{equation}
\cos(\Phi_\pm) = \mp\frac{L^2-kr}{r\sqrt{2EL^2+k^2}}=\cos(\phi_0),
\quad
\sin(\Phi_\pm) = \pm\frac{Lv}{\sqrt{2EL^2+k^2}}
\end{equation}
with $2EL^2+k^2>0$ for $E>E_\min$. 
Hence the vector first integral \eqref{ndim-Thetavec-alt} is given by 
\begin{equation}\label{kelper-negE-Thvec}
\Thvec_\pm 
=\frac{\mp 1}{r^2\sqrt{2EL^2+k^2}}\left( (L^2-kr)\rvec - (\vvec\cdot\rvec)\rvec\cdot\Ltens \right)
=\frac{\pm 1}{\sqrt{E|\Ltens|^2+k^2}}\left( (k/|\rvec|)\rvec +\vvec\cdot\Ltens \right) .
\end{equation}
Then the expression \eqref{ndim-LRLvec} for the general \LRL/ vector gives
\begin{equation}\label{kelper-negE-LRLvec}
\LRLvec_{*\pm} = \pm \left( (k/|\rvec|)\rvec +\vvec\cdot\Ltens \right)
\end{equation}
where we have again chosen $A(E,L)=\sqrt{E|\Ltens|^2+k^2}$ 
as the normalization factor. 
This vector \eqref{kelper-negE-LRLvec}
lies along the semi-major axis of the trajectory,
such that $\LRLvec_{*-}$ points toward the periapsis point 
and $\LRLvec_{*+}$ points toward the apoapsis point. 

Notice the relationship $\LRLvec_{*-}= \LRLvec_*$. 
This means that the expression \eqref{kelper-posE-E=0-LRLvec} 
provides a general \LRL/ vector that points toward the periapsis point 
for all three types of trajectories. 
In $n=3$ dimensions, this vector is exactly the usual \LRL/ vector \cite{GolPooSaf,Cor}
\begin{equation}
\vec A_* = \vvec\times\vec L - k|\rvec|^{-1}\rvec
\end{equation}
where $|\vec A| = \sqrt{2EL^2+k^2}$. 

An interesting variant of the $n$-dimensional general \LRL/ vector \eqref{kelper-posE-E=0-LRLvec} 
is given by choosing the radial value $r_0$ to be an inertial point on a trajectory. 
All three types of trajectories each have two inertial points \eqref{kepler-ip}
differing by $\sgn(v)\gtrless 0$ on the trajectory. 
Using the radius of the inertial points for the radial value $r_0$ 
in evaluating the angular expression \eqref{Theta0}, we have 
\begin{equation}
\Phi=\arctan\left(\frac{L^2-kr}{Lvr}\right)
\end{equation}
which yields 
\begin{equation}
\cos(\Phi) = \frac{|L||v|}{\sqrt{2EL^2+k^2}}=\cos(\phi_0),
\quad
\sin(\Phi) = \sgn(Lv) \frac{L^2-kr}{r\sqrt{2EL^2+k^2}}
\end{equation}
The vector first integral \eqref{ndim-Thetavec} is then given by 
\begin{equation}\label{kepler-ip-Thvec}
\begin{aligned}
\Thvec & 
=\frac{|L|}{r^2\sqrt{2EL^2+k^2}}\left( r|v|\rvec+\sgn(v)(1-kr/L^2)\rvec\cdot\Ltens \right)
\\&
= \frac{1}{\sqrt{2(E+k^2/|\Ltens|^2)}}\frac{1}{|\vvec\cdot\rvec|}
\left( (2E+k/|\rvec|)\rvec +(1-2k|\rvec|/|\Ltens|^2)\vvec\cdot\Ltens \right) . 
\end{aligned}
\end{equation}
Hence the expression \eqref{ndim-LRLvec} for the general \LRL/ vector gives
\begin{equation}\label{kepler-ip-LRLvec}
\LRLvec^* = \frac{1}{|\vvec\cdot\rvec|}
\left( (2E+k/|\rvec|)\rvec +(1-2k|\rvec|/|\Ltens|^2)\vvec\cdot\Ltens \right)
\end{equation}
where we have chosen $A(E,L)=\sqrt{2(E+k^2/|\Ltens|^2)}$.
(Note the superscript $*$ indicates the use of an inertial point for $r_0$
in defining the vector.)
This vector \eqref{kepler-ip-LRLvec} is related to the previous vector $\LRLvec_*$ by
\begin{equation}
\LRLvec_*\cdot\Lhattens = \sgn(v) \LRLvec^*
\end{equation}
which represents a rotation of $\sgn(\vvec\cdot\rvec)\pi/2$ in the plane of motion. 

Thus, the variant \LRL/ vector \eqref{kepler-ip-LRLvec} lies along a line that is perpendicular to the \LRL/ vector in the plane of motion. 
(In the case of elliptic trajectories, this line is the semi-minor axis.)
Most interestingly, 
it is double-valued since it changes sign around the periapsis point on a trajectory. 
In $n=3$ dimensions, this vector is a multiple of Hamilton's eccentricity vector \cite{Cor}
\begin{equation}
\vec e = \vvec + k|\vec L|^{-2}|\rvec|^{-1}\rvec\times\vec L 
= |\vec L|^{-2} \vec L\times\LRLvec_* = \sgn(v)|\vec L|^{-1} \LRLvec^*
\end{equation}
Although the eccentricity vector $\vec e$ is usually defined only for elliptic trajectories, 
the variant vector $\LRLvec^*$ exists for parabolic and hyperbolic trajectories as well.

\subsection{Inverse-square force with cubic corrections}
For an inverse-square force with cubic corrections \eqref{invsqcorrF}, 
there are again three different types of non-radial trajectories, 
depending on the first integrals \eqref{ndim-Ltens} and \eqref{ndim-E}
for angular momentum $\Ltens$ and energy $E$:
bounded precessing elliptic-like trajectories, with $0>E>E_\min$;
unbounded parabolic-like trajectories, with $E=0$;
and unbounded hyperbolic-like trajectories, with $E>0$.

Bounded trajectories have turning points \eqref{newton-negE-tp} 
given by the periapsis and apoapsis points on the trajectory. 
Using the radius of the turning points for the radial value $r_0$ 
to evaluate the angular expression \eqref{Theta0}, we have 
\begin{equation}
\Phi_\pm= \frac{L}{\sqrt{L^2-\kappa}}\Big( \arctan\left(\frac{L^2-\kappa -kr}{rv\sqrt{L^2-\kappa}}\right) \pm\sgn(v)\frac{\pi}{2} \Big)
\end{equation} 
where $L=\tfrac{1}{2}\Ltens\cdot\Lhattens$ is the angular momentum scalar 
and $v=\vvec\cdot\rhatvec$ is the radial speed. 
This yields 
\begin{equation}
\begin{aligned}
& 
\cos(\Phi_\pm) = \mp\cos\Big( \frac{L}{\sqrt{L^2-\kappa}}\arccos\Big(\frac{L^2-\kappa-kr}{r\sqrt{2E(L^2-\kappa)+k^2}}\Big) \Big)
=\cos(\phi_0),
\\
& 
\sin(\Phi_\pm) = \pm\sin\Big( \frac{L}{\sqrt{L^2-\kappa}}\arcsin\Big(\frac{v\sqrt{L^2-\kappa}}{\sqrt{2E(L^2-\kappa)+k^2}}\Big) \Big) . 
\end{aligned}
\end{equation}
In terms of these expressions, 
the vector first integral \eqref{ndim-Theta} is given by 
\begin{equation}\label{newton-negE-Thvec}
\begin{aligned}
\Thvec_\pm & = 
\mp\bigg( \cos\Big( \frac{L}{\sqrt{L^2-\kappa}}\arccos\Big(\frac{L^2-\kappa-kr}{r\sqrt{2E(L^2-\kappa)+k^2}}\Big) \Big)\rhatvec 
\\&\qquad
-\sin\Big( \frac{L}{\sqrt{L^2-\kappa}}\arcsin\Big(\frac{v\sqrt{L^2-\kappa}}{\sqrt{2E(L^2-\kappa)+k^2}}\Big) \Big) \rhatvec\cdot\Lhattens \bigg) . 
\end{aligned}
\end{equation}
The alternative formulation \eqref{ndim-Thetavec-alt} of this first integral 
is most easily obtained by re-writing the two terms in the expression \eqref{newton-negE-Thvec} by the following steps. 
First, we use the relations and \eqref{ndim-rvec-vvec} and \eqref{Lhattens-rvec-thvec} 
to express the second term in the form 
\begin{equation}\label{newton-negE-term2}
\sin(\Phi_\pm) \rhatvec\cdot\Lhattens 
= \frac{L}{rv}\sin(\Phi_\pm) \rhatvec +\frac{1}{v}\sin(\Phi_\pm) \vvec\cdot\Lhattens . 
\end{equation}
Next, we combine the coefficients of the $\rhatvec$ terms in expressions 
\eqref{newton-negE-Thvec} and \eqref{newton-negE-term2} to get
\begin{equation}\label{newton-negE-rterms}
\cos(\Phi_\pm)+\frac{L}{rv}\sin(\Phi_\pm) = 
\frac{\sqrt{r^2v^2+L^2}}{rv}\cos(\Phi_\pm-\Upsilon)
\end{equation}
where 
\begin{equation}\label{newton-Upsilon}
\Upsilon=\arctan\Big(\frac{L}{rv}\Big) . 
\end{equation}
We then obtain
\begin{equation}
\Thvec_\pm = \frac{1}{v}\Big( \frac{\sqrt{r^2v^2+L^2}}{r^2}\cos(\Phi_\pm -\Upsilon)\rvec +\frac{1}{L}\sin(\Phi_\pm) \vvec\cdot\Ltens \Big)
\end{equation}
which can be written out in an explicit form 
with the use of the relation \eqref{newton-v}. 
The expression \eqref{ndim-LRLvec} for the general \LRL/ vector thereby yields
\begin{equation}\label{newton-negE-LRLvec}
\begin{aligned}
\LRLvec_{*\pm} & = 
\frac{A(E,L)}{\vvec\cdot\rvec}\bigg( \sqrt{2(E+k/|\rvec|)+\kappa/|\rvec|^2}\times
\\&\qquad
\cos\Big( \frac{L}{\sqrt{L^2-\kappa}}\arccos\Big(\frac{\mp(L^2-\kappa-k)|\rvec|}{r\sqrt{2E(L^2-\kappa)+k^2}}\Big) 
-\arccos\Big(\frac{\vvec\cdot\rvec}{\sqrt{2(E+k/|\rvec|)+\kappa/|\rvec|^2}}\Big) \Big)\rvec 
\\&\qquad
\pm L^{-1}|\rvec|\sin\Big( \frac{L}{\sqrt{L^2-\kappa}}\arcsin\Big(\frac{\vvec\cdot\rvec\sqrt{L^2-\kappa}}{|\rvec|\sqrt{2E(L^2-\kappa)+k^2}}\Big) \Big)\vvec\cdot\Ltens \bigg)
\end{aligned}
\end{equation}
in terms of the energy $E$ and the scalar angular momentum $L$. 
(Note the subscript $*$ indicates the use of a turning point for $r_0$
in defining the vector.)
One possible choice for the normalization factor, 
in analogy to the Kepler case, is 
$A(E,L)=\sqrt{2E(L^2-\kappa)+k^2}$. 
The properties of this vector \eqref{newton-negE-LRLvec} are, however, 
quite different compared to the Kepler case. 
Since unbounded trajectories are precessing, 
$\LRLvec_{*\pm}$ is multi-valued such that, on a trajectory $\rvec(t)$, 
$\LRLvec_{*-}$ points toward the periapsis point closest to $\rvec(t)$ 
and $\LRLvec_{*+}$ points toward the apoapsis point closest to $\rvec(t)$. 
This means $\LRLvec_{*\pm}$ is jump discontinuous when $\rvec(t)$ passes 
through each periapsis and apoapsis, respectively. 

Unbounded trajectories have a single turning point 
\eqref{newton-E=0-tp} in the parabolic-like case 
and \eqref{newton-posE-tp} in the hyperbolic-like case. 
In both cases, the turning point is the periapsis on the trajectory. 
Evaluating the angular expression \eqref{Theta0} 
by using the radius $r_{*\pm}$ of the periapsis for the radial value $r_0$, 
we obtain $\Phi=\Phi_-$. 
Hence the vector first integral \eqref{ndim-Thetavec-alt} is given by 
$\Thvec= \Thvec_-$. 
Then $\LRLvec= \LRLvec_-$ is the expression for the general \LRL/ vector \eqref{ndim-LRLvec}. 
This vector lies along the radial line connecting the origin to the periapsis point on the trajectory. 

The expression 
\begin{equation}\label{newton-LRLvec}
\begin{aligned}
\LRLvec_* & = 
\frac{A(E,L)}{\vvec\cdot\rvec}\bigg( \sqrt{2(E+k/|\rvec|)+\kappa/|\rvec|^2}\times
\\&\qquad
\cos\Big( \frac{L}{\sqrt{L^2-\kappa}}\arccos\Big(\frac{L^2-\kappa-k|\rvec|}{|\rvec|\sqrt{2E(L^2-\kappa)+k^2}}\Big) 
-\arccos\Big(\frac{\vvec\cdot\rvec}{\sqrt{2(E+k/|\rvec|)+\kappa/|\rvec|^2}}\Big) \Big)\rvec 
\\&\qquad
-L^{-1}|\rvec|\sin\Big( \frac{L}{\sqrt{L^2-\kappa}}\arcsin\Big(\frac{\vvec\cdot\rvec\sqrt{L^2-\kappa}}{|\rvec|\sqrt{2E(L^2-\kappa)+k^2}}\Big) \Big)\vvec\cdot\Ltens \bigg)
\end{aligned}
\end{equation}
thereby provides a general \LRL/ vector that points toward the periapsis point closest to $\rvec(t)$,  
for all three types of trajectories. 

Similarly to the Kepler case, 
there is a variant of the $n$-dimensional general \LRL/ vector \eqref{newton-LRLvec},
which arises from choosing the radial value $r_0$ to be an inertial point on a trajectory. 
All three types of trajectories each have two inertial points \eqref{newton-ip}
differing by $\sgn(v)\gtrless 0$ on the each part trajectory. 
Using the radius of the inertial points for the radial value $r_0$ 
in evaluating the angular expression \eqref{Theta0}, we have 
\begin{equation}
\Phi=\frac{L}{\sqrt{L^2-\kappa}} \arctan\left(\frac{L^2-\kappa -kr}{rv\sqrt{L^2-\kappa}}\right) 
\end{equation}
which yields 
\begin{equation}
\begin{aligned}
& 
\cos(\Phi) = \cos\Big( \frac{L}{\sqrt{L^2-\kappa}}\arccos\Big(\frac{|v|\sqrt{L^2-\kappa}}{\sqrt{2E(L^2-\kappa)+k^2}}\Big) \Big)
=\cos(\phi_0),
\\
& 
\sin(\Phi) = \sgn(v)\sin\Big( \frac{L}{\sqrt{L^2-\kappa}}\arcsin\Big(\frac{L^2-\kappa-kr}{r\sqrt{2E(L^2-\kappa)+k^2}}\Big) \Big) . 
\end{aligned}
\end{equation}
The vector first integral \eqref{ndim-Thetavec} is then given by 
\begin{equation}\label{newton-ip-Thvec}
\begin{aligned}
\Thvec_\pm & = 
\cos\Big( \frac{L}{\sqrt{L^2-\kappa}}\arccos\Big(\frac{|v|\sqrt{L^2-\kappa}}{\sqrt{2E(L^2-\kappa)+k^2}}\Big) \Big) \rhatvec 
\\&\qquad
+\sgn(v)\sin\Big( \frac{L}{\sqrt{L^2-\kappa}}\arcsin\Big(\frac{L^2-\kappa-kr}{r\sqrt{2E(L^2-\kappa)+k^2}}\Big) \Big) \rhatvec\cdot\Lhattens . 
\end{aligned}
\end{equation}
By the same steps as in Kepler case, this leads to the expression 
\begin{equation}\label{newton-ip-LRLvec}
\begin{aligned}
\LRLvec^* & = 
\frac{A(E,L)}{\vvec\cdot\rvec}\bigg( \sqrt{2(E+k/|\rvec|)+\kappa/|\rvec|^2}\times
\\&\qquad
\cos\Big( \frac{L}{\sqrt{L^2-\kappa}}\arccos\Big(\frac{|\vvec\cdot\rvec|\sqrt{L^2-\kappa}}{|\rvec|\sqrt{2E(L^2-\kappa)+k^2}}\Big) 
-\arccos\Big(\frac{\vvec\cdot\rvec}{\sqrt{2(E+k/|\rvec|)+\kappa/|\rvec|^2}}\Big) \Big)\rvec 
\\&\qquad
+\sgn(v)L^{-1}|\rvec|\sin\Big( \frac{L}{\sqrt{L^2-\kappa}}\arcsin\Big(\frac{
L^2-\kappa-k|\rvec|}{|\rvec|\sqrt{2E(L^2-\kappa)+k^2}}\Big) \Big)\vvec\cdot\Ltens \bigg) .
\end{aligned}
\end{equation}
(Note the superscript $*$ indicates the use of an inertial point for $r_0$
in defining the vector.)
The variant vector $\LRLvec^*$ is related to the previous vector $\LRLvec_*$ 
by a rotation through an angle 
$\sgn({\vvec\cdot\rvec})(L/\sqrt{L^2-\kappa})\pi/2$ in the plane of motion. 
Hence, in contrast to the general \LRL/ vector $\LRLvec_*$, 
this variant vector $\LRLvec^*$ changes sign when $\rvec(t)$ passes 
through each periapsis on a trajectory.

\section{Concluding remarks}
\label{remarks}

The results presented in \secref{derivation} and \secref{ndim} provide 
a simple direct derivation of all $2n$ first integrals for general central force dynamics in $n>1$ dimensions. 
This derivation is based on solving the determining equations for first integrals of the \eom/ in a polar formulation 
and does not involve any use of symmetry. 
The first integrals are shown to be generated by 
an antisymmetric tensor (bi-vector) $\Lhattens$ determining the plane of motion,
the angular momentum $L$ in this plane, and the energy $E$,
which are defined for all solutions $\rvec(t)$, 
plus an angular quantity $\Theta$ and a temporal quantity $T$ 
both of which are defined only for non-circular solutions $\rvec(t)$. 
The quantities $\Lhattens$, $L$, $E$ comprise $2n-2$ functionally independent first integrals,
which constitute a complete set for all circular solutions, 
while the quantities $\Theta$, $T$ together with $\Lhattens$, $L$, $E$ 
comprise $2n$ functionally independent first integrals,
which constitute a complete set for all non-circular solutions. 

The angular quantity $\Theta$ is used to define a general \LRL/ vector
whose geometrical and physical properties are discussed in detail in \secref{ndim}. 
In particular, 
$\Theta$ is shown to be single-valued if a trajectory $\rvec(t)$ 
has at most one apsis (turning point) in the plane of motion. 
In this case the general \LRL/ vector has a unique direction 
aligned with the apsis on the trajectory. 
If instead a trajectory has multiple apses in the plane of motion
then $\Theta$ is shown to be single-valued when it is evaluated 
between pairs of successive periapsis points or apoapsis points on the trajectory. 
In this case, 
$\Theta$ defines a unique direction only if the angular separation between
any pair of successive periapsis points or apoapsis points on the trajectory
is an integer multiple of $2\pi$,
corresponding to the trajectory being closed. 
If the angular separation is not an integer multiple of $2\pi$, 
then the direction defined by $\Theta$ undergoes a discontinuous jump
when the next successive periapsis or apoapsis point is reached on the trajectory,
corresponding to the trajectory being either open or precessing. 

A variant of the general \LRL/ vector is also defined in \secref{ndim}, 
by using a different form of the angular quantity $\Theta$ 
based on the inertial points of a trajectory $\rvec(t)$. 
In this case, $\Theta$ is single-valued only on each piece of a trajectory 
where the radial speed has a definite sign,
while at each apsis point on the trajectory, $\Theta$ has a discontinuous jump. 
This variant is a generalization of Hamilton's eccentricity vector, 
which applies to general central forces. 

These properties are explicitly illustrated in \secref{examples} and \secref{LRLexamples}
for an inverse-square central force, 
where bounded trajectories are closed and non-precessing, 
and for a inverse-cube corrections to an inverse-square central force,
where bounded trajectories are precessing and either open or closed
(depending on the correction parameter). 

A symmetry interpretation of the first integrals $\Theta$ and $T$ is also obtained, using the Lagrangian formulation of the central force \eom/. 
In contrast to the well-known origin of $L$ and $E$ from point symmetries 
given by polar rotations and time translations, 
the quantities $\Theta$ and $T$ are shown to arise from hidden dynamical (first-order) symmetries. 
The explicit transformations generated by these symmetries are derived
by utilizing the action of the symmetries on the first integrals $L$, $E$, $\Theta$, $T$. 
These transformations are shown to form a four-dimensional abelian Lie group. 
This leads to a novel method for directly deriving the first integrals 
by use of extended point symmetries outlined in \secref{noether}. 

All of these new results will be further developed in future work. 
We plan to apply the method of extended point symmetries to give 
a new symmetry derivation of the $2n$ first integrals for the central force \eom/ in $n$ dimensions
and also to obtain the explicit hidden symmetry group underlying the general $n$-dimensional \LRL/ vector.

\section*{Acknowledgements}

S.C.A.\ is supported by an NSERC research grant. 
Georgios Papadopoulos is thanked for stimulating discussions on this work.

\end{document}